\documentclass[prb,aps,preprintnumbers,twocolumn, acknowledgement]{revtex4}

\usepackage{amsmath,amssymb,bm}
\usepackage{graphicx}
\usepackage{color}

\def\eqa{\begin{eqnarray}}
\def\eea{\end{eqnarray}}
\newcommand{\eq}{\begin{equation}}
\newcommand{\ee}{\end{equation}}

\newcommand{\Tr}{{\rm Tr}}

\begin{document}
\title{Anomalous Hall effect in type-I Weyl metals beyond the noncrossing approximation}
\author{Jia-Xing Zhang$^{1}$}
\author{Wei Chen$^{1,2}$} \email{chenweiphy@nju.edu.cn}
\affiliation{$^{1}$National Laboratory of Solid State Microstructures and School of Physics, Nanjing University, Nanjing, China}
\affiliation{$^2$Collaborative Innovation Center of Advanced Microstructures, Nanjing University, Nanjing, China}

\begin{abstract}
We study the anomalous Hall effect (AHE) in tilted Weyl metals with Gaussian disorder due to the crossed $X$ and $\Psi$ diagrams in this work. The importance of such diagrams to the AHE has been demonstrated recently in two dimensional (2D) massive Dirac model and Rashba ferromagnets. It has been shown that the inclusion of such diagrams dramatically changes the total AHE in such systems. In this work, we show that the contributions from the $X$ and $\Psi$ diagrams to the AHE in tilted Weyl metals are of the same order of the non-crossing diagram we studied in a previous work, but with opposite sign.  The total contribution of the $X$ and $\Psi$ diagrams cancels the majority part of the contribution from the non-crossing diagram in tilted Weyl metals, similar to the 2D massive Dirac model. We also discuss the difference of the contributions from the crossed diagrams between 2D massive Dirac model and the  tilted Weyl metals.
At last, we discuss the experimental relevance of observing the AHE due to the $X$ and $\Psi$ diagrams in  type-I Weyl metal such as ${\rm Co_3Sn_2S_2}$.
\end{abstract}

\maketitle

\section{Introduction}\label{Sec:I}
The anomalous Hall effect (AHE) has been a topic of  interest since it is first observed in ferromagnetic iron by Edwin Hall in 1881~\cite{Hall1881}. It is analogous to a usual Hall effect but without the need of an external magnetic field~\cite{Luttinger1954, Haldane1988}. The transverse motion in the anomalous Hall systems originates from the spin-orbit interaction and to have a net transverse current, the time reversal symmetry (TRS) has to be broken in the system~\cite{Luttinger1954, Haldane1988, Sinitsyn2007, Sinitsyn2008, Yang2011}. In insulator or semiconductor, the anomalous Hall conductivity is quantized and insensitive to impurity scatterings. In metals, however, the impurity scatterings affect the AHE significantly, 
and the AHE in such cases can be divided to the intrinsic contribution, which is due to the non-trivial topology of the electronic band structure and remains in the clean limit, and the extrinsic contribution, which is due to the impurity scatterings~\cite{Sinitsyn2007}. The anomalous Hall current can be obtained either from the quantum Kubo-Streda (QKS) formula~\cite{Streda1982} or from a semi-classical Boltzmann equation (SBE) approach~\cite{Sinitsyn2007, Sinitsyn2008}.  The former approach is more systematic whereas the latter is physically more transparent.

The extrinsic AHE depends on the type of impurities in general. For simplicity, we focus on the Gaussian white noise disorder in this work. It has been well-known that the Feynman diagrams with crossed impurity lines shown in Fig.\ref{fig:diagrams}(b), (c) and (d) result in a longitudinal conductivity which is smaller than the non-crossing diagram by a factor of $1/\epsilon_F \tau$ due to the restricted phase space of two rare impurity scatterings, so the crossed diagrams are usually neglected in computing the longitudinal conductivity. For a long time, the crossed diagrams were also ignored for the AHE and only the Feynman diagrams with non-crossing impurity scattering lines are considered for the AHE~\cite{Sinitsyn2007, MacDonald2006, Sinitsyn2008}. 
However, for the AHE, both the non-crossing and crossed diagrams contain the rare impurity processes~\cite{Levchenko2016, Ado2016} and are suppressed by $1/k_F l \ll1$ compared to the longitudinal conductivity for weak impurity systems. 
The two types of diagrams may then contribute the same order of magnitude to the AHE
as was demonstrated  in recent years in 2D Rashba ferromagnets and 2D massive Dirac model~\cite{Ado2015, Ado2016, Ado2017}.

The account of the  crossed diagrams in a number of  anomalous Hall systems changes the total AHE in the systems dramatically~\cite{Ado2016, Ado2015, Ado2017, Levchenko2016}. For example, the inclusion of the X and $\Psi$ diagrams in Fig.\ref{fig:diagrams}(b), (c) and (d)  in 2D Rashba ferromagnetic metal results in a non-vanishing AHE instead of the vanishing result under the non-crossing approximation (NCA)~\cite{Inoue2006, Sinova2007, Ado2016}. In the 2D massive Dirac model, the X and $\Psi$ diagrams almost cancel out the NCA contribution at high energy~\cite{Ado2015, Ado2017}. It was also shown that the same crossed diagrams play an important role for the AHE on the surface of topological Kondo insulator~\cite{Levchenko2016}, for the Kerr effect in chiral p-wave superconductors~\cite{Levchenko2017}, and for the extrinsic spin Hall effect across the weak and strong scattering regimes~\cite{Milletari2016, Ferreira2016}.

The above cases show that the crossed diagrams play an important role for a more complete study of the AHE in a general case. 
For that reason, we study the contributions of the crossed diagrams, namely the X and $\Psi$ diagrams to the AHE in three dimensional (3D) tilted Weyl metals with breaking TRS~\cite{Wan2011, Vishwanath2018} and weak Gaussian disorder  in this work. 
Although the SBE approach can yield the same result for the AHE as the QKS formula for the non-crossing diagram in isotropic systems~\cite{Sinitsyn2007}, it is not convenient for calculating the AHE from the $X$ and $\Psi$ diagrams because it is very difficult to compute the  scattering rates for such diagrams. We then employ the QKS formula and the diagrammatic technique to study the $X$ and $\Psi$ diagrams in tilted Weyl metals in this work.
Diagrams with more crossed impurity lines have smaller contribution in $1/\epsilon_F \tau$  for Gaussian disorder.  

For  untilted Weyl metals it has been shown that the impurity scatterings have little effect on the AHE only if the Fermi energy is not very far  from the Weyl nodes~\cite{Burkov2014}. This is because the low energy effective Hamiltonian of a single Weyl node of the untilted Weyl metal gains an emergent TRS and the AHE due to the  impurity scatterings vanishes. For tilted Weyl metals, the tilting breaks the TRS of the effective Hamiltonian of a single Weyl node~\cite{Pesin2017, Zyuzin2017} and the impurity scatterings have significant effects on the AHE in such system~\cite{Fu2021, Chen2022}. 
In a previous paper~\cite{Chen2022}, we have studied the disorder induced AHE in the tilted Weyl metals due to the non-crossing diagrams and obtained both the intrinsic and extrinsic contribution for such diagrams from the quantum Kubo-Streda formula. We also separated the two different extrinsic contributions, namely the side jump and skew scattering contribution from the non-crossing diagrams in this system. The study of the crossed diagrams for the tilted Weyl metals in this work is an important supplement of the skew scattering contribution to the AHE in such system. 

The skew scattering contribution to the AHE comes from the diffractive  scattering off two  impurities, as can be seen from the crossed $X$ and $\Psi$ diagrams~\cite{Levchenko2016},  as well as the non-crossing skew scattering diagrams in Ref.~\cite{Sinitsyn2007}.  For the two scattering processes to interfere, the two impurities need to be close enough for Gaussian disorder, i.e., with distance of the order of the Fermi wavelength. This is verified by the calculation of the AHE from the $X$ and $\Psi$ diagrams in the real space~\cite{Ado2016, Levchenko2016}.  

We show that the  contribution from both the $X$ and $\Psi$ diagrams for tilted Weyl metals with Gaussian disorder is of the same order of the contribution from the NCA diagram we studied in the previous work~\cite{Chen2022}, i.e., $\sim$$\tau^0$. 
This is different from the 2D massive Dirac model for which the contribution from the $\Psi$ diagram vanishes for Gaussian disorder~\cite{Ado2015, Ado2017}.   On the other hand, our calculation shows that the total contribution of the $X$ and $\Psi$ diagram cancels a majority part of the contribution from the NCA diagram for tilted Weyl metals. This is similar to the 2D massive Dirac model. However, the inclusion of the $X$ and $\Psi$ diagram
does not change the dependence of the anomalous Hall conductivity on the Fermi energy whereas in 2D massive Dirac model, the crossed diagram changes the total anomalous Hall conductivity from $\sigma_{xy}\sim m/\epsilon_F$ for NCA diagram to $\sigma_{xy}\sim (m/\epsilon_F)^3$~\cite{Ado2017}.

We also discussed the experimental relevance of observing the effects of the $X$ and  $\Psi$ diagrams in tilted Weyl metals, such as ${\rm Co_3Sn_2S_2}$~\cite{Felser2018, Ding2019, Wang2018}. We point out that the density of the Gaussian disorder needed to observe the contributions of the crossed $X$ and $\Psi$ diagrams is much higher than that of  observing the non-crossing diagram with single impurity scatterings, such as the side jump contribution, since the former corresponds to electron scatterings by pairs of closely located impurities with distance of the order of the Fermi wavelength. We estimated that the  impurity density needed to observe the AHE due to the $X$ and $\Psi$ diagrams is $n^{sk}_{\rm imp}>\sqrt{2}/(\lambda_F l_\phi)^{3/2}$ for 3D tilted Weyl metals, where $l_\phi$ is the phase coherence length and $\lambda_F$ is the Fermi wavelength and $l_\phi\gg \lambda_F$  (see the Discussion section). As a comparison, the impurity density needed to observe the AHE from the non-crossing diagram with single impurity scatterings is $n^{sj}_{\rm imp}>1/(l_\phi)^{3}$ which is much lower than $n^{sk}_{\rm imp}$.
Another issue is that the intrinsic AHE from the Chern-Simons term is much higher than the AHE from both the non-crossing and crossed diagrams in ${\rm Co_3Sn_2S_2}$ with Gaussian disorder so the effects of the Gaussian disorder on the AHE is not very distinguishable in experiments in this system. We propose that one can observe the effects of the Gaussian disorder on the AHE by measuring the anomalous Nernst effect (ANE) in such a system because the Chern-Simons term has no contribution to the ANE and the contributions of the Gaussian disorder to the ANE and AHE are proportional to each other. 

This paper is organized as follows. In Sec.~II, we present the model and the calculation of the anomalous Hall effect due to the crossed $X$ and $\Psi$ diagrams in tilted Weyl metals, and 
 compare the AHE from the crossed diagrams  with the non-crossing diagram, as well as with other systems, such as 2D massive Dirac model. 
In Sec.~III, we have a discussion of the experimental relevance of observing the effects of $X$ and $\Psi$ diagrams. In Sec.~IV, we have a summary of this work.

\section{AHE in tilted Weyl metals due to X and $\Psi$ diagrams}
The low energy physics of a Type-I Weyl metal with breaking TRS can be described by an effective low energy Hamiltonian of two independent Weyl nodes and a topological Chern-Simons  term~\cite{Burkov2015, Chen2022}. The Chern-Simons term results in an AHE proportional to the distance of the two Weyl nodes and is not affected by the impurity scatterings. We will then focus on the low energy effective Hamiltonian of the Weyl nodes in the following, which is 
\begin{equation}\label{eq:tilted_Hamiltonian}
H=\sum_\chi (\chi v \boldsymbol \sigma\cdot \mathbf{p}+\mathbf {u_\chi \cdot p}),
\end{equation}
where $\chi=\pm 1$ is the chirality of the two Weyl nodes, $\boldsymbol \sigma$ are the Pauli matrices and $\mathbf{u}_\chi$ is a tilting velocity with $u_\chi<v$ for type-I Weyl metals we consider in this work.  Here we assume the tilting $\mathbf{u}_+=-\mathbf{u}_-=\mathbf{u}$, i.e., the tilting is  opposite for the two valleys. For this case, the AHE in the two Weyl nodes adds up instead of cancel out. 
The Hamiltonian $H_\chi$ for each single valley results in two tilting  linear bands $\epsilon_{\pm}= \pm v p + \mathbf {u_\chi \cdot p}$. The tilting term   breaks the TRS of a single Weyl node so the AHE from each valley does not vanish. The term  $\chi v \mathbf{\boldsymbol \sigma\cdot p}$ breaks the global TRS so the total AHE of the two valleys is non-zero.

\begin{figure}
	\includegraphics[width=8cm]{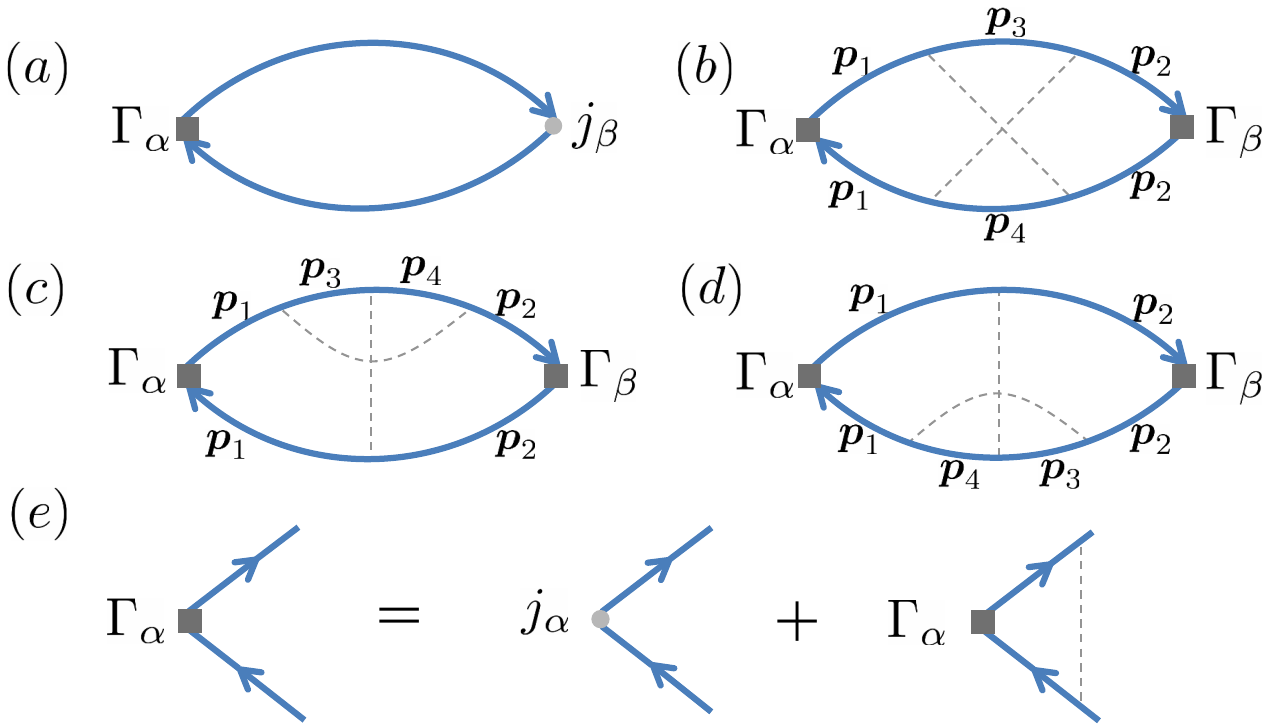}
		\caption{(a)The Feynman diagram of the response function $\Pi^{\mathbf{I}}_{\alpha\beta}$ under the non-crossing approximation (NCA) in the spin basis. The thick solid lines are Green's function in the spin basis under the Born approximation, and the solid square represents the current vertex $\Gamma_\alpha$ renormalized by the ladder diagram under the NCA. (b)The X diagram with two crossed impurity lines. (c) and (d): The $\Psi$ diagram with two crossed impurity lines. (e)The recursion equation satisfied by the renormalized current vertex $\Gamma_\alpha$.}\label{fig:diagrams} 
\end{figure}

We consider  weak Gaussian disorder (white noise) with random potential $V({\mathbf r})=V_0 \sum_a \delta({\mathbf r-r_a})$ and correlation $\langle V({\mathbf r})V({\mathbf r'}) \rangle=\gamma \delta({\mathbf r}-{\mathbf r'})$, where $\gamma=n_{\rm imp} V_0^2$ and $n_{\rm imp}$ is the impurity density. We assume that all the higher order correlators of the impurity potential vanish for simplicity, and  the mean free path of the electrons is much larger than the Fermi wavelength, i.e., $k_F l \gg 1$ or $\epsilon_F\tau\gg 1$.  The anomalous Hall conductivity may be written in two parts, i.e., $\sigma_H^{\rm I}$ and $\sigma_H^{\rm II}$, in the Kubo-Streda formula\cite{Streda1982, Sinitsyn2007}. Formally $\sigma_H^{\rm I}$ takes into account the contribution on the Fermi surface, and  $\sigma_H^{\rm II}$ includes contribution from the whole Fermi sea. Since $\sigma_H^{\rm II}$ is not sensitive to impurity scatterings and its contribution in the clean limit has been studied in previous works for tilted Weyl metals\cite{Pesin2017, Zyuzin2017},  we only need to study the $\sigma_H^{\rm I}$ part in this work. 
The leading order contribution to the response function $\Pi^{\mathbf{I}}_{\alpha\beta}$ includes the diagrams  in Fig.\ref{fig:diagrams}, where Fig.\ref{fig:diagrams}a is the diagram under the NCA and has been studied in our previous work~\cite{Chen2022}. The NCA diagram includes both the intrinsic and extrinsic contributions, and both contributions to the AHE are independent of the scattering rate $1/\tau$ in the leading order, i.e.,  $\sim\tau^0$. For the crossed diagrams in Fig. \ref{fig:diagrams}(b)-(d),  previous works~\cite{Ado2016, Ado2017, Levchenko2016} have shown that for Gaussian disorder, the leading order AHE from  these  diagrams for 2D Rashba ferromagnets and massive Dirac models is of the same order as the non-crossing diagram in Fig.\ref{fig:diagrams}a, , i.e., $\sim \tau^0$.
 In the following, we study the contribution of the crossed $X$ and $\Psi$ diagrams  for tilted Weyl metals and compare the leading order contribution of these diagrams with the non-crossing diagram in Fig.\ref{fig:diagrams}a. Diagrams with more crossed impurity lines have smaller contribution in $1/\tau$ and so are negligible.

 We assume that the impurity potential is diagonal for both the spin and valley index, so the two valleys decouple and one can compute the AHE in each valley separately. 
 The leading order contribution to the AHE from the NCA diagram of the tilted Weyl metals has been worked out in our previous work~\cite{Chen2022}. The total dc anomalous Hall conductivity from the non-crossing diagram for the two Weyl nodes  is  $\sigma^{\rm NCA}_{xy}=4e^2 \epsilon_F u/3\pi^2 v^2$ in the leading order of $u/v$ for $\mathbf u$ in the $z$ direction. As a comparison, we compute the  anomalous Hall conductivity in the dc limit due to the crossed $X$ and $\Psi$ diagram in the tilted Weyl metals  in the leading order of $u/v$  in the following. 

We consider a uniform electric field ${\mathbf E}=-\partial_t {\mathbf A}$ applied to the system. In the linear response regime $j^{\rm \mathbf{I}}_\alpha=\Pi^{\rm \mathbf{I}}_{\alpha\beta} A^\beta$, where $A^\alpha=(0, \mathbf{A})$.
The response functions $\Pi^{\mathbf{I}}_{\alpha\beta}$ in the dc limit for the X and $\Psi$ diagrams  for a single Weyl node (e.g., $\chi=1$)   are respectively 
\begin{eqnarray}
&& \Pi_{\alpha\beta}^{X}
=\gamma^2\omega\sum_{ \mathbf{p}_1,..., \mathbf{p}_4}                                                                                                   
   \int \frac{d\epsilon}{2\pi i}  \frac{d n_F(\epsilon)}{d\epsilon}
   \delta( \mathbf{p}_1+ \mathbf{p}_2- \mathbf{p}_3- \mathbf{p}_4)\nonumber\\
 &&\rm Tr [{\Gamma}_{\alpha}G^R(\mathbf{p}_1)G^R(\mathbf p_3)G^R(\mathbf p_2){\Gamma}_{\beta}G^A(\mathbf p_2)G^A(\mathbf p_4)G^A(\mathbf p_1)], \label{eq:X_diagram}\nonumber\\
  \end{eqnarray}
  and
 \begin{eqnarray}
  &&  \Pi_{\alpha\beta}^{\Psi}
  =\gamma^2\omega\sum_{\mathbf{p}_1, ..., \mathbf{p}_4}
   \int \frac{d\epsilon}{2\pi i}  \frac{d n_F(\epsilon)}{d\epsilon}
   \delta(\mathbf p_1+\mathbf p_4-\mathbf p_2-\mathbf p_3)\nonumber\\
&&\Tr [G^A(\mathbf p_1){\Gamma}_{\alpha}G^R(\mathbf p_1)G^R(\mathbf p_3)
        G^R(\mathbf p_4)G^R(\mathbf p_2){\Gamma}_{\beta}G^A(\mathbf p_2)\nonumber\\
  &&+G^A(\mathbf p_1){\Gamma}_{\alpha}G^R(\mathbf p_1)
       G^R(\mathbf p_2){\Gamma}_{\beta}G^A(\mathbf p_2)
       G^A(\mathbf p_4)G^A(\mathbf p_3)], \label{eq:Psi_diagram}\nonumber\\
\end{eqnarray}
where $G^{R/A}$ is the retarded/advanced Green's function (GF) of the tilted Weyl metals, and $\Gamma_\alpha$ is the  current vertex renormalized by the non-crossing ladder diagram~\cite{Chen2022}. We have omitted the argument $\epsilon$ in $G^{R/A}(\epsilon, \mathbf p)$ in the above equations for brevity.

The impurity averaged retarded/advanced GF in a single valley (e.g. with $\chi=1$) under the first Born approximation is ~\cite{Chen2022}
\begin{equation}\label{eq:GF}
G^{R/A}(\epsilon, \mathbf{p})=(\epsilon- v \mathbf{\boldsymbol\sigma\cdot p}-\mathbf{u\cdot p} -\Sigma^{R/A})^{-1},
\end{equation}
 where the self-energy  due to the impurity scatterings is $\Sigma^{R/A}=\gamma\sum_{\mathbf p}G^{R/A}_0({\mathbf p})=\mp\frac{i}{2\tau}[1+\mathbf{\Delta}(\mathbf{u})\cdot {\boldsymbol \sigma}]$ with $1/\tau=\pi \gamma g(\epsilon_F)$,  
 $g(\epsilon_F)=\int \frac{d^3 p}{(2\pi)^3}\delta(\mathbf{u\cdot p} + v p -\epsilon_F)=\frac{\epsilon^2_F v}{2\pi^2(v^2-u^2)^2}$ being the density of states at the Fermi energy $\epsilon_F>0$ and ${\mathbf\Delta}(\mathbf{u})=-{\mathbf u}/v$.
We note here that the self-energy under the self-consistent Born approximation produces the same AHE as that under the first Born approximation in the leading order of $\epsilon_F\tau$ for the Gaussian disorder. The inclusion of diagrams with crossed impurity lines in the self-energy also only results in corrections to the AHE in the higher orders of $1/\epsilon_F\tau$.

For the calculation in this work, it is  convenient to write the $G^{R/A}$ in Eq.(\ref{eq:GF}) as
\begin{equation}
G^{R/A}(\epsilon, {\mathbf p})=\frac{(\epsilon\pm\frac{i}{2\tau}-\boldsymbol u \cdot \boldsymbol p)\sigma^0+v\boldsymbol p \cdot \boldsymbol \sigma \mp\frac{i}{2\tau}(\boldsymbol \Delta \cdot \boldsymbol \sigma)}{(\epsilon-\epsilon_p^{+}\pm\frac{i}{2\tau^{+}})(\epsilon-\epsilon_p^{-}\pm\frac{i}{2\tau^{-}})},
\end{equation}
with
$1/\tau^{\pm}=\frac{1}{\tau}(1\pm \frac{\mathbf p\cdot \mathbf{\Delta}}{p}).$

The renormalized current vertex $\Gamma_\alpha$ in Eq.(\ref{eq:X_diagram}) and (\ref{eq:Psi_diagram})
has been worked out in our previous work~\cite{Chen2022}. The bare current vertex for the tilted Weyl metals is $\hat{j}_\alpha=e(v\sigma_\alpha+u_\alpha\sigma_0)
$ (we define $u_0\equiv 0$). By expressing $\hat{j}_\alpha$ and $\Gamma_\alpha$ with the Pauli matrices as $\hat{j}_\alpha={\cal J}_{\alpha \beta} \sigma_\beta$ and $\hat{\Gamma}_\alpha=\Gamma_{\alpha \beta} \sigma_\beta, \alpha, \beta=0, x, y, z$, one can solve the coefficients of the renormalized current vertex as $\Gamma_{\alpha \beta} ={\cal J}_{\alpha \gamma}{\cal D}_{\gamma \beta} $, where the summation over the repeated index $\gamma$ is implied as usual and ${\cal D}=(1-\gamma{\cal I})^{-1}$  is the $4\times 4$ diffusion matrix with the polarization operator ${\cal I}$ defined as
\begin{equation}\label{eq:I_matrix_0}
{\cal I}_{\alpha\beta}=
\frac{1}{2}\int \frac{d{\mathbf p}}{(2\pi )^3}\rm{Tr} [\sigma_\alpha G^R(\epsilon+\omega, \mathbf{p}+\mathbf{q})  \sigma_\beta G^A(\epsilon, \mathbf{p})]. 
\end{equation}

In our previous work~\cite{Chen2022}, we have shown that the renormalized current vertex $\Gamma_{\alpha\beta}=ev {\cal D}_{\alpha\beta}$, i.e., the tilting term $u_\alpha \sigma_0$ in the bare current vertex has no contribution to the AHE and the main effect of  
the tilting is to produce an anisotropy of the Fermi surface. 
We have also worked out the  ${\cal I}$ matrix and ${\cal D}$ matrix for the tilted Weyl metals in the previous work~\cite{Chen2022}, so we will just apply the results of such matrices for the study in this work.

 Denoting the integrand of ${\cal I}_{\alpha\beta}$ in the dc limit as $I_{\alpha\beta}(\mathbf{p})=\frac{1}{2}\rm{Tr} [\sigma_\alpha G^R(\epsilon, \mathbf{p})  \sigma_\beta G^A(\epsilon, \mathbf{p})]$, one gets $G^A \hat{\Gamma}_\alpha G^R=\Gamma_{\alpha\beta}I_{\beta\gamma}\sigma_{\gamma},\ G^R \hat{\Gamma}_\alpha G^A=\sigma_\gamma I_{\gamma\beta}\Gamma_{\alpha\beta}$. 
The response functions for the X and $\Psi$ diagrams in Fig.\ref{fig:diagrams}(b)-(d) can then be written as
\begin{widetext}
\begin{eqnarray}
    \Pi_{\alpha\beta}^{X}
   &=&e^2\gamma^2 v^2\omega\int \frac{d\epsilon}{2\pi i}\frac{d n_F(\epsilon)}{d\epsilon}\sum_{\mathbf p_1,..., \mathbf p_4}\delta( \mathbf{p}_1+ \mathbf{p}_2- \mathbf{p}_3- \mathbf{p}_4) {\cal D}_{\alpha\xi}I_{\xi \mu}(\mathbf p_1)F_{\mu\nu}(\mathbf p_3, \mathbf{p}_4)I_{\nu\gamma}(\mathbf p_2){\cal D}^T_{\gamma\beta}, \label{eq:X_diagram_1} \\
    \Pi_{\alpha\beta}^{\Psi}
  &=&e^2\gamma^2 v^2\omega\int \frac{d\epsilon}{2\pi i}\frac{d n_F(\epsilon)}{d\epsilon}\sum_{\mathbf p_1,..., \mathbf p_4}\delta( \mathbf{p}_1+ \mathbf{p}_4- \mathbf{p}_2- \mathbf{p}_3) {\cal D}_{\alpha\xi}I_{\xi \mu}(\mathbf p_1)M_{\mu\nu}(\mathbf p_3,\mathbf  p_4)I_{\nu\gamma}(\mathbf p_2){\cal D}^T_{\gamma\beta}, \label{eq:Psi_diagram_1}
    \end{eqnarray}
    \end{widetext}
where we have defined 
\begin{eqnarray}
    F_{\mu\nu}(\mathbf p_3,\mathbf p_4)&\equiv& \Tr[\sigma_{\mu}G^R(\epsilon, \mathbf p_3)\sigma_{\nu}G^A(\epsilon, \mathbf p_4)], \label{eq:f_function}\\
    M_{\mu \nu}(\mathbf p_3,\mathbf p_4)&\equiv&\Tr[\sigma_{\mu}\sigma_{\nu}G^{A}(\epsilon, \mathbf p_4)G^A(\epsilon, \mathbf p_3) \nonumber\\
    &&\ \ \  +\sigma_{\nu}\sigma_{\mu}G^{R}(\epsilon, \mathbf p_3)G^R(\epsilon, \mathbf p_4)].  \label{eq:M_function}
\end{eqnarray}

The AHE due to the X and $\Psi$ diagrams corresponds to the anti-symmetric part of the response function $\Pi_{\alpha\beta}^{X}$ and $\Pi_{\alpha\beta}^{\Psi}$. In the following, we study the AHE in tilted Weyl metals due to the two diagrams respectively.

\subsection{AHE from the X diagram}

In this subsection, we study the AHE due to the $X$ diagram in tilted Weyl metals. To do this, we first compute the anti-symmetric part of the response function $\Pi^X_{\alpha\beta}$ in Eq.(\ref{eq:X_diagram_1}).

For the matrices ${\cal D}, I$ and $F$  in Eq.(\ref{eq:X_diagram_1}), the symmetric parts of these matrices are ${\cal D}_0\sim \tau^0, I^s\sim \tau, F^s\sim\tau^0$, and the anti-symmetric parts ${\cal D}^a\sim \tau^{-1}, I^a\sim \tau^0, F^a\sim \tau^0$. 
In the leading order of $1/\epsilon_F \tau$, the anti-symmetric part of $\Pi_{\alpha\beta}^{X}$ is then 
\begin{eqnarray}\label{eq:X_diagram_2}
   && \Pi_{\alpha\beta}^{X, a}
   =e^2\gamma^2\omega\int \frac{d\epsilon}{2\pi i}\frac{d n_F(\epsilon)}{d\epsilon}\sum_{\mathbf p_1,\mathbf p_2, \mathbf Q}\nonumber\\
 && {\cal D}_{0, \alpha\gamma}I^s_{\gamma \mu}(\mathbf p_1)F^a_{\mu\nu}(\mathbf{p}_1-\mathbf Q, \mathbf{p}_2+\mathbf Q)I^s_{\nu\eta}(\mathbf p_2){\cal D}^T_{0, \eta\beta}, \nonumber\\
\end{eqnarray}
where $\mathbf Q\equiv \mathbf p_1-\mathbf p_3=\mathbf p_4-\mathbf p_2$.

The vertex correction factor ${\cal D}_0$ and ${\cal D}^T_0$ on the two ends of $\Pi^{X, a}_{\alpha\beta}$ are constant matrices as a function of $u$ and $v$, as given in our previous work~\cite{Chen2022},  and when multiplied with the remaining part of the response function, it  results in an extra total factor $\tilde{\alpha}^2\approx 9/4+{\cal O}(u^2/v^2)$ only if the remaining part is an anti-symmetric matrix of the linear order of $u_i, i=1, 2, 3$, which is the case for both the X and $\Psi$ diagrams. For convenience, we will then drop the ${\cal D}_0$ factor in Eq.(\ref{eq:X_diagram_2}) in the following calculation and add the vertex correction factor $\tilde{\alpha}^2$ at the end.

Since the symmetric part of the $I$ matrix in the dc limit is 
\begin{equation}\label{eq:I_s}
I^s_{\alpha\beta}(\mathbf p)\approx \pi \tau^+ \delta(\epsilon-\boldsymbol u \cdot \boldsymbol p-vp)\frac{1}{p^2}\times  p_{\alpha}p_{\beta},
\end{equation}
the integration over the momentum $\mathbf{p}_1$ and $\mathbf{p}_2$ is  bound to the Fermi surface due to the $\delta$ function in $I^s$.

The anti-symmetric part of the $F$ matrix for the X diagram is $F^a_{\mu\nu}=N_{\mu\nu}(F^a)/D(F)$, where
\begin{eqnarray} 
 &N_{\mu\nu}(F^a)= 2iv^2 \{ \epsilon^{0\mu\nu k}[(p_1p_{2k}-p_2p_{1k})+(p_1+p_2)Q_k \nonumber\\
&\ \ \ \ \ \ \ \ +\frac{\mathbf u \cdot \mathbf Q}{v}(p_{1k}+p_{2k})] -\epsilon^{\mu\nu l k}(\mathbf p_1-\mathbf Q)_l(\mathbf p_2+\mathbf Q)_k \}, \nonumber\\
 &D(F)
=[\epsilon-\mathbf{u \cdot}(\mathbf{p}_1-\mathbf{Q})-v|\mathbf{p}_1-\mathbf{Q}|+\frac{i}{2\tau_3^+}] \nonumber\\
 & \ \ \ \ \ \ \ \ \times [\epsilon-\mathbf{u \cdot}(\mathbf{p}_1-\mathbf{Q})+v|\mathbf{p}_1-\mathbf{Q}|+\frac{i}{2\tau_3^-}] \nonumber\\
&   \ \ \ \ \ \ \ \ 
  \times[\epsilon-\mathbf{u \cdot}(\mathbf{p}_2+\mathbf{Q})-v|\mathbf{p}_2+\mathbf{Q}|-\frac{i}{2\tau_4^+}] \nonumber\\
& \ \ \ \ \ \ \ \ \ \times [\epsilon-\mathbf{u \cdot}(\mathbf{p}_2+\mathbf{Q})+v|\mathbf{p}_2+\mathbf{Q}|-\frac{i}{2\tau_4^-}],
\end{eqnarray}
and $\frac{1}{\tau_i^{\pm}}\equiv\frac{1}{\tau}(1\pm \delta_i),\delta_i\equiv\frac{\mathbf{p}_i\cdot\mathbf{\Delta}}{p_i}$,  $\mu, \nu=0, 1, 2, 3$, and $l, k=1, 2, 3$.  In $N_{\mu\nu}(F^a)$, we have only kept the leading order in $1/\tau$.

We assume $\mathbf u$ in the $z$ direction for simplicity and $\boldsymbol Q=Q(\sin \alpha \cos \beta,\sin \alpha \sin \beta,\cos \alpha)$. Rotate the $z$-axis to the direction of $\boldsymbol Q$ by the transformation
\begin{displaymath}
  \left(
  \begin{array}{c}
    \hat{\mathbf x'}\\
    \hat{\mathbf  y'} \\
    \hat{\mathbf  z'}
  \end{array}\right)=\left(
  \begin{array}{ccc}
    \cos \alpha \cos \beta & \cos \alpha \sin \beta & -\sin \alpha \\
    -\sin \beta & \cos \beta & 0 \\
    \sin \alpha \cos \beta & \sin \alpha \sin \beta & \cos \alpha
  \end{array}\right) \
  \left( \begin{array}{c}
   \hat{\mathbf  x} \\
   \hat{\mathbf  y} \\
    \hat{\mathbf  z}
         \end{array}\right),
\end{displaymath}
where $(\hat{\mathbf x}, \hat{\mathbf y}, \hat{\mathbf z})$ and $(\mathbf x', \mathbf y', \mathbf z')$ are the bases of the old and new frames respectively. The coordinates of $\mathbf p_i, i=1, 2$ in the old and new frames are denoted as $(p_{ix}, p_{iy}, p_{iz})$ and $(p'_{ix}, p'_{iy}, p'_{iz})$
respectively. 
Assuming in the rotated frame $\mathbf p_i=p\ (\sin \theta_i \cos \phi_i \cdot \hat{\mathbf x'} +\sin \theta_i \sin \phi_i \cdot \hat{\mathbf y'}+\cos \theta_i \cdot \hat{\mathbf z'})$ for  $i=1, 2$, we then have $\mathbf{p}_1 \cdot \mathbf{Q}=p_1Q\cos\theta_1,\mathbf{p}_2 \cdot \mathbf{Q}=p_2Q\cos\theta_2, \mathbf u \cdot \mathbf Q=u Q \cos \alpha$. 

 The coordinates $p_{i\alpha}, i=1, 2$ in the old frame may be expressed as 
\begin{widetext}
\begin{eqnarray}
\left\{\begin{array}{ll}
p_{i, x}=&p_i(\cos \phi_i \sin \theta_i\cos \alpha \cos \beta- \sin \theta_i \sin \phi_i\sin\beta +\cos \theta_i \sin \alpha\cos\beta), \\
p_{i, y}=&p_i (\cos \phi_i \sin \theta_i\cos \alpha\sin \beta+ \sin \theta_i \sin \phi_i\cos\beta +\cos \theta_i \sin \alpha\sin\beta),\\
p_{i, z}=&p_i(-\sin \theta_i \cos \phi \sin \alpha+\cos \theta_i \cos \alpha).\\
                     \end{array}\right.
\end{eqnarray}

From the $\delta$ function in $I^s$, one can get $p_i=\frac{\epsilon}{v+u\hat{z}_i}, i=1, 2$, where 
\begin{eqnarray}
\hat{z}_i&=&p_{iz}/p_i= -\sin \theta_i \cos \phi_i \sin \alpha+\cos \theta_i \cos \alpha. 
\end{eqnarray}

Applying the $\delta$ function in $I^s$ to replace $\epsilon$ by $\mathbf p_1$ and $\mathbf p_2$ in $D(F)$,  we get 
\begin{eqnarray}
\frac{1}{D(F)}=
&[(v^2-u^2\cos^2\alpha)Q^2-2v^2p_1Q(\cos\theta_1+\frac{u}{v}\cos\alpha)-\frac{i}{\tau}(vp_1+uQ\cos\alpha+v\delta_3|\mathbf{p}_1-\mathbf{Q}|)]^{-1}
\nonumber\\
&\times[(v^2-u^2\cos^2\alpha)Q^2+2v^2p_2Q(\cos\theta_2+\frac{u}{v}\cos\alpha)+\frac{i}{\tau}(vp_2-uQ\cos\alpha+v\delta_4|\mathbf{p}_2+\mathbf{Q}|)]^{-1}.
\end{eqnarray}
\end{widetext}

The AHE due to the $X$ and $\Psi$ diagram is finite only when $\mathbf u$ is non-zero. In this work, we only compute the  AHE in the tilted Weyl metals in the leading order of $\mathbf u$ for simplicity. For the reason, we expand $1/D(F)$ in terms of $u/v$ and keep only the terms up to the  linear order of $u$. We then get 
\begin{widetext}
\begin{eqnarray}\label{eq:DF_expansion}
  \frac{1}{D(F)}
  &\approx& [v^2Q^2-2v^2 \boldsymbol p_1 \cdot \boldsymbol Q-\frac{i}{\tau}vp_1-2v p_1\boldsymbol u\cdot \boldsymbol Q]^{-1}
  [v^2Q^2+2v^2 \boldsymbol p_2 \cdot \boldsymbol Q+\frac{i}{\tau}vp_2+2v p_2\boldsymbol u\cdot \boldsymbol Q]^{-1}\nonumber\\
 &\approx &\frac{1+\frac{u}{v}\hat{z}_1}{v^2Q^2-2v\epsilon Q(\cos\theta_1+\frac{u}{v}\cos\alpha)-\frac{i}{\tau}\epsilon +vuQ^2\hat{z}_1}
  \frac{1+\frac{u}{v}\hat{z}_2}{v^2Q^2+2v\epsilon Q(\cos\theta_2+\frac{u}{v}\cos\alpha)+\frac{i}{\tau}\epsilon +vuQ^2\hat{z}_2}\nonumber\\
  & \approx& (1+\frac{u}{v}\hat{z}_1)(1+\frac{u}{v}\hat{z}_2)
  \frac{1}{v^2Q^2-2v\epsilon Q\cos\theta_1-\frac{i}{\tau}\epsilon}(1-\frac{vuQ^2\hat{z}_1-2\epsilon uQ\cos\alpha}{v^2Q^2-2v\epsilon Q\cos\theta_1-\frac{i}{\tau}\epsilon})\nonumber\\
  &&  \times \frac{1}{v^2Q^2+2v\epsilon Q\cos\theta_2+\frac{i}{\tau}\epsilon}(1-\frac{vuQ^2\hat{z}_2+2\epsilon uQ\cos\alpha}{v^2Q^2+2v\epsilon Q\cos\theta_2+\frac{i}{\tau}\epsilon}),
\end{eqnarray}
\end{widetext}
where we have neglected the linear order of $u$ terms $\sim i u Q \cos \alpha/\tau$ and $i \delta_3/\tau, i\delta_4/\tau$ since they contain an extra small factor $1/\tau$.

Putting $I^s$ and $F^a$ together and neglecting the vertex correction at the two ends of the $X$  diagram at the moment, we get the anti-symmetric part of the response function  $\Pi_{xy}^{X}$ as
\begin{widetext}
\begin{eqnarray}\label{eq:X_diagram_3}
  \Pi_{\alpha\beta}^{X, a}
   &=&\pi^2 e^2\gamma^2\omega  \int \frac{d\epsilon}{2\pi i}\frac{d n_F(\epsilon)}{d\epsilon}\int_0^{\infty} \frac{dp_1}{8\pi^3}p^2_1 \int_0^{\infty} \frac{dp_2}{8\pi^3} p^2_2 \int_{0}^{\infty} \frac{dQ}{8\pi^3} Q^2
   \int_{0}^{\pi}\sin\theta_1 d\theta_1 \int_{0}^{\pi}\sin\theta_2 d\theta_2 \int_{0}^{\pi}\sin\alpha d\alpha \nonumber\\
&&\int_{0}^{2\pi}d\phi_1 \int_{0}^{2\pi} d\phi_2 \int_{0}^{2\pi} d\beta\frac{4iv^2  p_{1\alpha}p_{2\beta}(p_1+p_2)(\boldsymbol p_1 \times \boldsymbol p_2)\cdot \boldsymbol Q }{p^2_1\ p^2_2\ D(F)}\underset{i=1,2}{\mathbf\Pi}\tau_i^+\delta(\epsilon-\boldsymbol u \cdot \boldsymbol p_i-vp_i).
\end{eqnarray}
\end{widetext}

The scalar factor $(\boldsymbol p_1 \times \boldsymbol p_2)\cdot \boldsymbol{Q}=p_1p_2Q\sin\theta_1\sin\theta_2\sin(\phi_2-\phi_1)$ in the rotated frame, and $\tau_i^+=\tau/(1-\frac{u}{v}\hat{z}_i), i=1, 2$. For the integrand in Eq.(\ref{eq:X_diagram_3}), only the factor $p_{1\alpha}p_{2\beta}$ includes the angle $\beta$ and one can easily integrate out this angle. For $\mathbf u$ in the $z$ direction, if the electric field $\mathbf E$ is also in the $z$ direction, $\Pi_{\alpha z}^{X, a}=0$ after the integration over the angle $\beta$ for $\alpha\neq z$.  For the reason, we only need to consider the case when $\mathbf E$ is perpendicular to $\mathbf u$. 
Assuming the electric field $\mathbf E$ in the $y$ direction, $\Pi_{zy}^{X, a}$ is zero with the integration over $\beta$. We then only need to compute the non-vanishing component $\Pi_{xy}^{X, a}$. 

Since
\begin{eqnarray}
&&\int_{0}^{2\pi} d\beta p_{1x}p_{2y} =\pi p_1p_2 [\cos\alpha \sin\theta_1\sin\theta_2\sin(\phi_2-\phi_1) \nonumber\\
&&\ \   +\sin\alpha (\cos\theta_1\sin\theta_2\sin \phi_2-\cos\theta_2\sin\theta_1\sin \phi_1)], 
\end{eqnarray}
the response function  $\Pi_{xy}^{X,a}$ for the $X$ diagram becomes
\begin{widetext}
\begin{eqnarray}\label{eq:Pi_X_4}
   \Pi_{xy}^{X,a}
   &=&e^2v^2\gamma^2\tau^2\omega \int \frac{d\epsilon}{2\pi i}\frac{d n_F(\epsilon)}{d \epsilon}
    \frac{1}{(2\pi)^9}4iv^2\pi^3 \nonumber\\
   &\times& \int_{0}^{\infty}Q^2 dQ \int_{0}^{\pi}\sin\alpha d\alpha
   \int_{0}^{\pi}\sin\theta_1 d\theta_1 \int_{0}^{\pi}\sin\theta_2 d\theta_2 \int_{0}^{2\pi}d\phi_1 \int_{0}^{2\pi} d\phi_2 \ Q\sin\theta_1\sin\theta_2\sin(\phi_2-\phi_1)\nonumber \\
   & \times& [\cos\alpha \sin\theta_1\sin\theta_2\sin(\phi_2-\phi_1) +\sin\alpha (\cos\theta_1\sin\theta_2\sin \phi_2-\cos\theta_2\sin\theta_1\sin \phi_1)]\nonumber\\
   &\times& \frac{v}{v-u\hat{z}_1}\frac{v}{v-u\hat{z}_2}(\frac{\epsilon}{v+u\hat{z}_1}+\frac{\epsilon}{v+u\hat{z}_2}) \frac{\epsilon^2}{(v+u\hat{z}_1)^3} \frac{\epsilon^2}{(v+u\hat{z}_2)^3} \frac{1}{D(F)}.
\end{eqnarray}

It is easy to check that at $u=0$, the response function in Eq.(\ref{eq:Pi_X_4}) vanishes. Expanding Eq.(\ref{eq:Pi_X_4}) to the linear order of $u$ and combing $1/D(F)$ in Eq.(\ref{eq:DF_expansion}) , we get 
\begin{eqnarray}
 \Pi_{xy}^{X, a}(u)
   &=&e^2v^2\gamma^2\tau^2\omega \int \frac{d\epsilon}{2\pi i}\frac{d n_F(\epsilon)}{d \epsilon}\times
   4iv^2\pi^3 \frac{1}{(2\pi)^9} \times \frac{\epsilon^5}{v^7}\nonumber\\
   &\times& \int_{0}^{\infty}Q^2 dQ \int_{0}^{\pi}\sin\alpha d\alpha
   \int_{0}^{\pi}\sin\theta_1 d\theta_1 \int_{0}^{\pi}\sin\theta_2 d\theta_2 \int_{0}^{2\pi}d\phi_1 \int_{0}^{2\pi} d\phi_2 \times Q\sin\theta_1\sin\theta_2\sin(\phi_2-\phi_1)\nonumber \\
   & \times& [\cos\alpha \sin\theta_1\sin\theta_2\sin(\phi_2-\phi_1) +\sin\alpha (\cos\theta_1\sin\theta_2\sin \phi_2-\cos\theta_2\sin\theta_1\sin \phi_1)]\nonumber\\
   &\times& \frac{1}{v^2Q^2-2v\epsilon Q\cos\theta_1-\frac{i}{\tau}\epsilon}
   \frac{1}{v^2Q^2+2v\epsilon Q\cos\theta_2+\frac{i}{\tau}\epsilon}\nonumber \\
   &\times& [-3\frac{u}{v}(\hat{z}_1+\hat{z}_2)-2\frac{vuQ^2\hat{z}_1-2\epsilon uQ\cos\alpha}{v^2Q^2-2v\epsilon Q\cos\theta_1-\frac{i}{\tau}\epsilon}-2\frac{vuQ^2\hat{z}_2+2\epsilon uQ\cos\alpha}{v^2Q^2+2v\epsilon Q\cos\theta_2+\frac{i}{\tau}\epsilon}].
\end{eqnarray}
\end{widetext}

The angular integration over  $\phi_1, \phi_2$ and $\alpha$ can be easily done in the above equation and the contributions 
from the terms with the factor $\hat{z}_1$ and $\hat{z}_2$ vanish after these  angular integration. The response function for the $X$ diagram after the integration over $\phi_1, \phi_2, \alpha$ and $\epsilon$ becomes 
\begin{widetext}
\begin{eqnarray}\label{eq:Pi_X_final}
 \Pi_{xy}^{X, a}(u)&=&-\frac{\omega}{12\pi^3}e^2v^3\epsilon_F^2 u \times \int_{0}^{\infty}Q^4 dQ
   \int_{0}^{\pi}\sin\theta_1 d\theta_1 \int_{0}^{\pi}\sin\theta_2 d\theta_2  \sin^2\theta_1\sin^2\theta_2 \frac{1}{v^2Q^2-2v\epsilon_F Q\cos\theta_1-\frac{i}{\tau}\epsilon_F}\nonumber\\
   &&\times [\frac{1}{v^2Q^2-2v\epsilon_F Q\cos\theta_1-\frac{i}{\tau}\epsilon_F}-\frac{1}{v^2Q^2+2v\epsilon_F Q\cos\theta_2+\frac{i}{\tau}\epsilon_F}] 
   \frac{1}{v^2Q^2+2v\epsilon_F Q\cos\theta_2+\frac{i}{\tau}\epsilon_F}\nonumber\\
&=&-i\omega\frac{u}{6\pi^3}e^2v^3\epsilon_F^2 \times \text{Im} \int_{0}^{\infty} dQ S^X(Q), \label{eq:Pi_X}
\end{eqnarray}
where 
\begin{equation}
S^X(Q)\equiv Q^4 \int_{0}^{\pi}\sin\theta_1 d\theta_1 \int_{0}^{\pi}\sin\theta_2 d\theta_2  \sin^2\theta_1\sin^2\theta_2
  \times\frac{1}{(v^2Q^2+2v\epsilon_F Q \cos\theta_1-\frac{i}{\tau}\epsilon_F)^2}\times \frac{1}{v^2Q^2+2v\epsilon_F Q \cos\theta_2+\frac{i}{\tau}\epsilon_F}.
\end{equation}
\end{widetext}

The anomalous Hall conductivity from the $X$ diagram is $\sigma^X_{xy}=\Pi^{X, a}_{xy}/i\omega$, which is then completely real and dissipationless. The integration in  Eq.(\ref{eq:Pi_X_final}) can be done by a change of variable $x=\cos \theta_1, y=\cos\theta_2$, as shown in the Appendix. We  get in the leading order of $1/\tau$
\begin{equation}\label{eq:I_X}
I^X\equiv {\rm Im} \int_0^{\infty} dQ\ S^{X}(Q)\approx  \frac{\pi}{\epsilon_F v^5}.
 \end{equation}
The leading order response function without vertex correction for the $X$ diagram  from a single valley  is
\begin{equation}
\Pi_{xy}^{X, a}(u)\approx -i\omega \frac{e^2\epsilon_F u}{6 \pi^2 v^2}.
\end{equation}

The vertex correction adds a factor of $9/4$ in the leading order of $u/v$ to the response function $\Pi_{xy}^{X}(u)$. For tilted Weyl metals with two valleys, the response function doubles. We then obtain the leading order anomalous Hall conductivity  of the tilted Weyl metals due to the $X$ diagram  as
\begin{equation}\label{eq:AHE_X}
\sigma^X_{xy}\approx - \frac{3e^2\epsilon_F u}{4\pi^2 v^2},
\end{equation}
which is of the same order of the leading order contribution from the NCA diagram  $\sigma^{\rm NCA}_{xy}=\frac{4e^2\epsilon_F u}{3\pi^2 v^2}$, but with opposite sign.

\subsection{AHE from the $\Psi$ diagram}

In this subsection, we study the AHE in tilted Weyl metals due to the $\Psi$ diagram. To do this, we  compute the anti-symmetric part of the response function $\Pi^{\Psi}_{\alpha\beta}$ in Eq.(\ref{eq:Psi_diagram_1}).

For the matrices ${\cal D}, I$ and $M$  in Eq.(\ref{eq:Psi_diagram_1}), the symmetric parts of these matrices are ${\cal D}_0\sim \tau^0, I^s\sim \tau,  M^s\sim \tau^0$, and the anti-symmetric parts ${\cal D}^a\sim \tau^{-1}, I^a\sim \tau^0, M^a\sim \tau^0$. 
In the leading order of $1/\epsilon_F \tau$, the anti-symmetric part of  $\Pi_{\alpha\beta}^{\Psi}$ is then 
\begin{widetext}
\begin{eqnarray}\label{eq:Psi_diagram_2}
 \Pi_{\alpha\beta}^{\Psi, a}
  =e^2\gamma^2\omega\int \frac{d\epsilon}{2\pi i}\frac{d n_F(\epsilon)}{d\epsilon}\sum_{\mathbf p_1,\mathbf p_2, \mathbf Q}
 {\cal D}_{0, \alpha\gamma}I^s_{\gamma \mu}(\mathbf p_1)M^a_{\mu\nu}(\mathbf Q-\mathbf{p}_2, \mathbf Q-\mathbf{p}_1)I^s_{\nu\eta}(\mathbf p_2){\cal D}^T_{0, \eta\beta}, 
\end{eqnarray}
where $\mathbf Q\equiv\mathbf p_1+\mathbf p_4=\mathbf p_2+\mathbf p_3$,  $I^s$ is given in Eq.(\ref{eq:I_s}) and $M^a_{\mu\nu}$ is the anti-symmetric part of 
\begin{equation}\label{eq:ker_Psi_2}
    M_{\mu \nu}\equiv\Tr[\sigma_{\mu}\sigma_{\nu}G^{A}(\mathbf p_4)G^A(\mathbf p_3)]
    +\Tr[\sigma_{\nu}\sigma_{\mu}G^{R}(\mathbf p_3)G^R(\mathbf p_4)].
\end{equation}

We denote the $G^A G^A$ term in Eq.(\ref{eq:ker_Psi_2}) as $M^A$ and the $G^RG^R$ term as $M^R$. 
Since $M^R=(M^A)^*$, the $M$ matrix is then $M=2\rm Re\ M^A$.
The anti-symmetric part of the $M^A$ matrix in the leading order of $1/\tau$ is $M^{A, a}_{\mu\nu}=N_{\mu\nu}(M^{A, a})/D(M^A)$, where
\begin{eqnarray}
N_{\mu\nu}(M^{A,a})& =& \ 2iv\epsilon^{0\mu\nu k}\{[2(\epsilon-\boldsymbol u\cdot \boldsymbol Q)+\boldsymbol u\cdot (\boldsymbol p_1+\boldsymbol p_2)]Q_k-[\epsilon-\boldsymbol u\cdot (\boldsymbol Q-\boldsymbol p_2)]p_{1k}-[\epsilon-\boldsymbol u\cdot (\boldsymbol Q-\boldsymbol p_1)]p_{2k}\} \nonumber\\
&&\ -2v^2[(Q-p_1)_{\mu}(Q-p_2)_{\nu}-(Q-p_1)_{\nu}(Q-p_2)_{\mu}], \label{eq:N_M}\\
 D(M^A)
     &=& [\epsilon-\boldsymbol u\cdot (\boldsymbol Q-\boldsymbol p_1)-v|\boldsymbol Q-\boldsymbol p_1|-\frac{i}{2\tau^+_4}]
    [\epsilon-\boldsymbol u\cdot (\boldsymbol Q-\boldsymbol p_1)+v|\boldsymbol Q-\boldsymbol p_1|-\frac{i}{2\tau^-_4}]\nonumber\\
    &\times&
    [\epsilon-\boldsymbol u\cdot (\boldsymbol Q-\boldsymbol p_2)-v|\boldsymbol Q-\boldsymbol p_2|-\frac{i}{2\tau^+_3}]
    [\epsilon-\boldsymbol u\cdot (\boldsymbol Q-\boldsymbol p_2)+v|\boldsymbol Q-\boldsymbol p_2|-\frac{i}{2\tau^-_3}].
\label{eq:D_M}
\end{eqnarray}
In the above equation, $\frac{1}{\tau_i^{\pm}}=\frac{1}{\tau}(1\pm \delta_i),\delta_i=\frac{\mathbf{p}_i\cdot\mathbf{\Delta}}{p_i}$ as defined before. 

For $\mathbf u$ in the $z$ direction and $\mathbf Q=Q(\sin\alpha \cos\beta, \sin\alpha \sin\beta, \cos \alpha)$, after the same rotation of the $z$-axis to the direction of $\mathbf Q$ as for the $X$ diagram, and applying the $\delta$ function in $I^s$, we obtain
\begin{equation}
D(M^A)\approx
 [v^2Q^2-2v^2 \mathbf{p_1\cdot Q}+\frac{i}{\tau}vp_1-2vp_1\boldsymbol u\cdot (2\boldsymbol p_1-\boldsymbol Q)] 
[v^2Q^2-2v^2\mathbf{p_2\cdot Q}+\frac{i}{\tau}vp_2-2vp_2\boldsymbol u\cdot (2\boldsymbol p_2-\boldsymbol Q)], 
\end{equation}
where we have neglected the second order $\mathbf u$ terms as well as the $\mathbf u/\tau$ terms.

Putting $I^s$ and $M^a$ together and neglecting the vertex correction at the two ends of the $\Psi$ diagram at the moment, we get the anti-symmetric part of the response function  for the $\Psi$ diagram as
\begin{eqnarray}\label{eq:Psi_diagram_3}
  \Pi_{\alpha\beta}^{\Psi, a} 
   &=&\pi^2 e^2\gamma^2\omega  \int \frac{d\epsilon}{2\pi i}\frac{d n_F(\epsilon)}{d\epsilon} \nonumber\\
  && \times\sum_{\mathbf p_1,\mathbf p_2, \mathbf Q}
\left[\frac{2iv\epsilon^{0\mu\nu k}p_{1, \alpha}p_{1, \mu}p_{2, \nu} p_{2, \beta}[2(\epsilon-\boldsymbol u\cdot \boldsymbol Q)+\boldsymbol u\cdot (\boldsymbol p_1+\boldsymbol p_2)]Q_k }{p^2_1\ p^2_2\ \rm  D(M^A)}+c.c\right]\underset{i=1,2}{\mathbf\Pi}\tau_i^+\delta(\epsilon-\boldsymbol u \cdot \boldsymbol p_i-vp_i)\nonumber\\
&=&e^2v^2\gamma^2\omega \int \frac{d\epsilon}{2\pi i}\frac{d n_F(\epsilon)}{d \epsilon}\times
   2iv\pi^2 \frac{1}{(2\pi)^9}\int_{0}^{\infty}Q^2 dQ \int_{0}^{\pi}\sin\alpha d\alpha \int_{0}^{2\pi}d\beta p_{1\alpha}p_{2\beta}\nonumber\\
   && \times \int_{0}^{\infty}dp_1 \int_{0}^{\infty} dp_2 \int_{0}^{\pi}\sin\theta_1 d\theta_1 \int_{0}^{\pi}\sin\theta_2 d\theta_2 \int_{0}^{2\pi}d\phi_1 \int_{0}^{2\pi} d\phi_2 \  \tau_1^+ \tau_2^+ (\boldsymbol p_1 \times \boldsymbol p_2)\cdot \boldsymbol{Q}\nonumber\\
  &&\times [2\epsilon+\mathbf{u}\cdot(\boldsymbol p_1+\boldsymbol p_2-2\boldsymbol Q) ] [\frac{1}{D(M^A)}-{\rm c.c}]\delta(\epsilon -vp_1-\boldsymbol u\cdot \boldsymbol p_1)
  \delta(\epsilon -vp_2-\boldsymbol u\cdot \boldsymbol p_2).
\end{eqnarray}

At $\mathbf u=0$,  the response function $\Pi_{\alpha\beta}^{\Psi, a}$ vanishes. We then expand $\Pi_{\alpha\beta}^{\Psi, a}$ to the linear order of $\mathbf u$ and neglect the higher order contributions. Similar to the $X$ diagram, for $\mathbf u$ in the $z$ direction and $\mathbf E$ in the $y$ direction, $\Pi^{\Psi, a}_{zy}=0$ and we only need to consider $\Pi^{\Psi, a}_{xy}$ for the AHE. Keeping only the linear order of $u$ and integrating out the angle $\beta$, we get
\begin{align}\label{eq:first order expansion of u/v in Pi^Psi}
   \Pi_{xy}^{\Psi, a}(u)
   &=e^2v^2\gamma^2\tau^2\omega \int \frac{d\epsilon}{2\pi i}\frac{d n_F(\epsilon)}{d \epsilon}
   4iv\pi^3 \frac{1}{(2\pi)^9}\  \frac{\epsilon^5}{v^6}\nonumber\\
   &\times \int_{0}^{\infty}Q^2 dQ \int_{0}^{\pi}\sin\alpha d\alpha
   \int_{0}^{\pi}\sin\theta_1 d\theta_1 \int_{0}^{\pi}\sin\theta_2 d\theta_2 \int_{0}^{2\pi}d\phi_1 \int_{0}^{2\pi} d\phi_2 \ Q\sin\theta_1\sin\theta_2\sin(\phi_2-\phi_1)\nonumber \\
   & \times [\cos\alpha \sin\theta_1\sin\theta_2\sin(\phi_2-\phi_1) +\sin\alpha (\cos\theta_1\sin\theta_2\sin \phi_2-\cos\theta_2\sin\theta_1\sin \phi_1)] \nonumber\\
  &\times \{\frac{1}{v^2Q^2-2v\epsilon Q\cos\theta_1+\frac{i}{\tau}\epsilon}
   \frac{1}{v^2Q^2-2v\epsilon Q\cos\theta_2+\frac{i}{\tau}\epsilon}\nonumber  [-\frac{u}{\epsilon}Q\cos\alpha + \frac{u}{2v} (\hat{z}_1+\hat{z}_2) \\
  & -\frac{2uv\hat{z}_1(Q^2-\frac{\epsilon}{v}Q\cos\theta_1-2\frac{\epsilon^2}{v^2})+2\epsilon uQ\cos\alpha}{v^2Q^2-2\epsilon vQ \cos\theta_1+\frac{i}{\tau}\epsilon} - \frac{2uv\hat{z}_2(Q^2-\frac{\epsilon}{v}Q\cos\theta_2-2\frac{\epsilon^2}{v^2})+2\epsilon uQ\cos\alpha}{v^2Q^2-2\epsilon vQ \cos\theta_2+\frac{i}{\tau}\epsilon}] -{\rm c.c}\}.
\end{align}

 After the integration over  $\phi_1, \phi_2$ and $\alpha$, we get the anti-symmetric part of $\Pi_{xy}^{\Psi}$ as
 \begin{eqnarray}
  \Pi_{xy}^{\Psi, a}(u)&=&-\frac{\omega}{12\pi^3}e^2v^3\epsilon_F^2 u  \int_{0}^{\infty}Q^4 dQ
   \int_{0}^{\pi}\sin\theta_1 d\theta_1 \int_{0}^{\pi}\sin\theta_2 d\theta_2  \sin^2\theta_1\sin^2\theta_2 \nonumber\\
   &&\times \frac{1}{2}\{ \frac{1}{v^2Q^2-2v\epsilon_F Q\cos\theta_1+\frac{i}{\tau}\epsilon_F}
   \frac{1}{v^2Q^2-2v\epsilon_F Q\cos\theta_2+\frac{i}{\tau}\epsilon_F} \nonumber\\
   &&\times [-\frac{1}{2\epsilon_F^2}-(\frac{1}{v^2Q^2-2\epsilon_F vQ \cos\theta_1+\frac{i}{\tau}\epsilon_F}
  + \frac{1}{v^2Q^2-2\epsilon_F vQ \cos\theta_2+\frac{i}{\tau}\epsilon_F})]-{\rm c.c}\}\nonumber\\
   &=& -i\omega\frac{u}{6\pi^3}e^2v^3\epsilon_F^2 \ \rm Im \int_0^{\infty} dQ S^{\Psi}(Q), \label{eq:Pi_Psi}
\end{eqnarray}
for which we separate $S^\Psi(Q)$ to two parts as 
\begin{eqnarray}
S^{\Psi}(Q)&=&S^{\Psi,1}(Q)+S^{\Psi,2}(Q), \nonumber\\
S^{\Psi,1}(Q)
    &=&-\frac{Q^4}{4\epsilon_F^2} \int_{0}^{\pi}\sin^3\theta_1 d\theta_1 \int_{0}^{\pi}\sin^3\theta_2 d\theta_2  
 \frac{1}{v^2Q^2-2v\epsilon_F Q\cos\theta_1+\frac{i}{\tau}\epsilon_F}
   \frac{1}{v^2Q^2-2v\epsilon_F Q\cos\theta_2+\frac{i}{\tau}\epsilon_F}, \label{eq:S_Psi_1}\\
S^{\Psi,2}(Q)   &=&
    -Q^4 \int_{0}^{\pi}\sin^3\theta_1 d\theta_1 \int_{0}^{\pi}\sin^3\theta_2 d\theta_2 
    \frac{1}{(v^2Q^2-2v\epsilon_F Q\cos\theta_1+\frac{i}{\tau}\epsilon_F)^2}
   \frac{1}{v^2Q^2-2v\epsilon_F Q\cos\theta_2+\frac{i}{\tau}\epsilon_F}. \label{eq:S_Psi_2}
\end{eqnarray}
\end{widetext}

As shown in the Appendix, in the leading order of $1/\tau$, the integration over $S^{\Psi,1}(Q)$ and $S^{\Psi,2}(Q)$ gives
\begin{equation}\label{eq:S_Psi1}
I^{\Psi, 1}\equiv {\rm Im} \int_0^{\infty} dQ\ S^{\Psi, 1}(Q)\approx  \frac{17+16  \ln 2}{105}  \frac{\pi}{\epsilon_F v^5}, 
 \end{equation}
 and 
 \begin{equation}\label{eq:S_Psi2}
I^{\Psi, 2}\equiv {\rm Im} \int_0^{\infty} dQ\ S^{\Psi, 2}(Q) \approx  \frac{1+8  \ln 2}{15}   \frac{\pi}{\epsilon_F v^5}.
 \end{equation}
 
The antisymmetric part of the response function for the $\Psi$ diagram  for a single valley  without vertex correction is 
\begin{equation}
 \Pi_{xy}^{\Psi, a}(u)=\Pi_{xy}^{\Psi, 1, a}(u)+\Pi_{xy}^{\Psi, 2, a}(u) 
 \approx  -\frac{4 +12 \ln 2}{105} i\omega \frac{e^2\epsilon_F u}{\pi^2 v^2}.
 \end{equation}
 
 Adding the vertex correction factor $9/4$ in the leading order of $u$, and taking into account the two valleys of the tilted Weyl metals, we get the total
 anomalous Hall conductivity due to the  $\Psi$ diagram in the leading order of $u$ as
 \begin{equation}
 \sigma^{\Psi}_{xy}\approx -\frac{6+18\ln 2}{35} \frac{e^2\epsilon_F u}{ \pi^2 v^2}\approx -0.53 \frac{e^2\epsilon_F u}{ \pi^2 v^2}.
 \end{equation}
 The contribution from the $\Psi$ diagram is  also of the same order of the contribution from the NCA diagram  $\sigma^{\rm NCA}_{xy}=\frac{4e^2\epsilon_F u}{3\pi^2 v^2}$, but with opposite sign. This is different from the 2D massive Dirac model, for which the anomalous Hall conductivity from the $\Psi$ diagram vanishes for Gaussian disorder~\cite{Ado2017}.

 \subsection{Comparison  between the crossed and non-crossing diagrams}
 
 \begin{figure}
	\includegraphics[width=8cm]{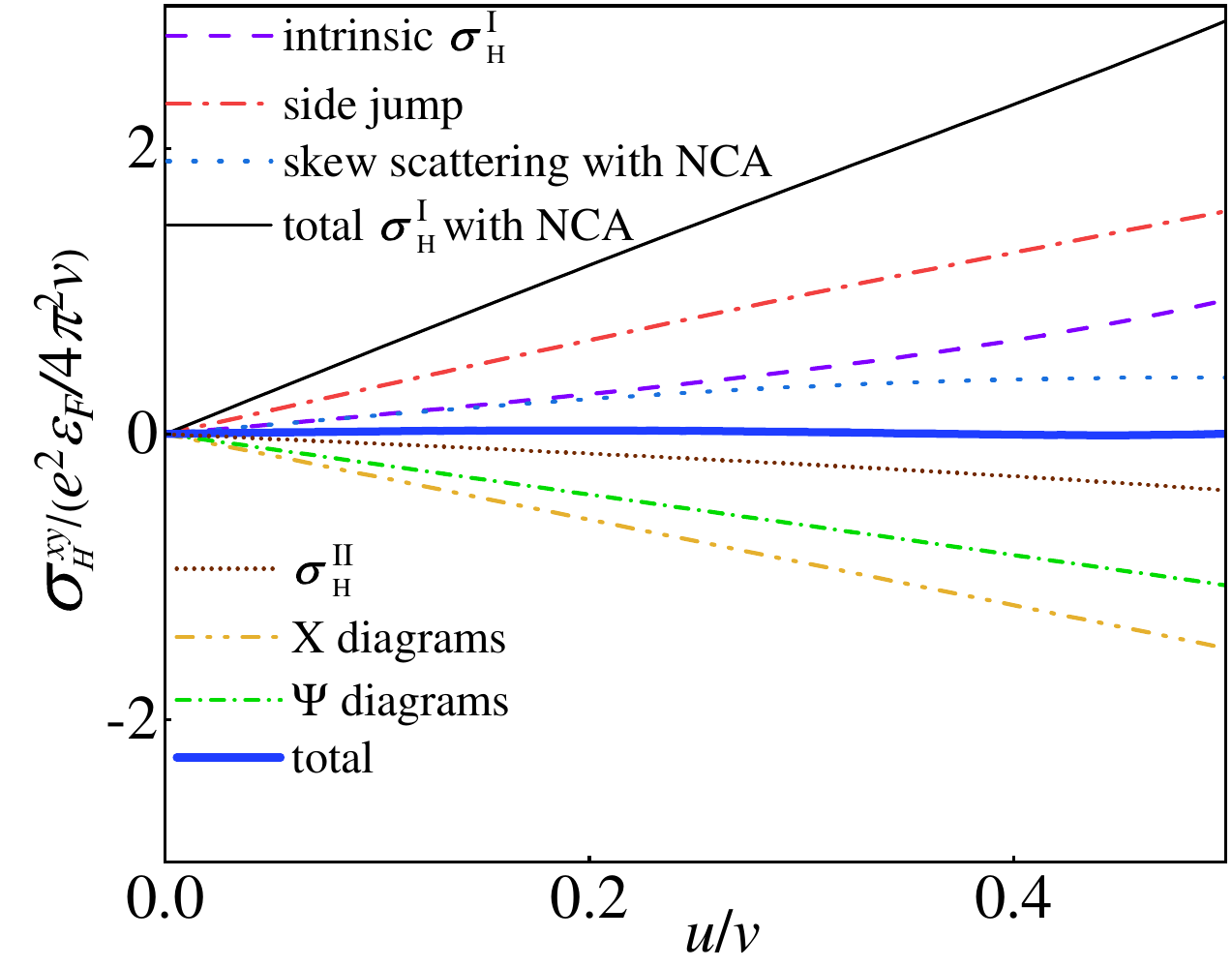}
		\caption{The different contributions to the anomalous Hall conductivity  from the non-crossing and crossed diagrams for the 3D tilted Weyl metals with Gaussian disorder as a function of the tilting velocity $u$ rescaled by $v$. The  Fermi surface contributions from the non-crossing diagram in the plot come from Ref.~\cite{Chen2022} and are exact, whereas the results for the crossed diagrams are kept in the linear order of $u/v$. The intrinsic contribution from the Fermi sea $\sigma^{\rm II}_H$ comes from Ref.\cite{Pesin2017}. The black solid line represents the total Fermi surface contribution from the non-crossing diagram, and the blue solid line represents the total contribution from both the Fermi surface and Fermi sea, including both non-crossing and crossed diagrams.}\label{fig:AH_conductivity} 
\end{figure}

The total contribution of the $X$ and $\Psi$ diagrams to the AHE for the  tilted Weyl metals with two valleys is  
\begin{equation}\label{eq:AHE_crossed}
 \sigma^{X+\Psi}_{xy}=\sigma^{X}_{xy}+\sigma^{\Psi}_{xy}\approx -1.28\frac{e^2\epsilon_F u}{ \pi^2 v^2}.
 \end{equation}

As a comparison, we plot the different contributions to the anomalous Hall conductivity  of the tilted Weyl metals due to both the non-crossing and crossed diagrams in Fig.\ref{fig:AH_conductivity}. The anomalous Hall conductivity from the non-crossing diagram was obtained in our previous work~\cite{Chen2022} and includes three different mechanisms: intrinsic, side jump and skew scattering. 
The total anomalous Hall conductivity from the non-crossing diagram for Gaussian disorder in the leading order of $u/v$ is 
\begin{equation}\label{eq:AHE_NCA}
 \sigma^{\rm NCA}_{xy}\approx\frac{4e^2\epsilon_F u}{ 3\pi^2 v^2}.
 \end{equation}
 This contribution includes the intrinsic part $\sigma^{\rm I}_{int}=\frac{e^2\epsilon_F u}{ 3\pi^2 v^2}$ from the Fermi surface and $\sigma^{\rm II}_{int}=-\frac{e^2\epsilon_F u}{ 6\pi^2 v^2}$ from the Fermi sea. The remaining part is the extrinsic contribution due to impurity scatterings, including the side jump and skew scattering contribution as shown in Fig.\ref{fig:AH_conductivity}.

From Eq.(\ref{eq:AHE_crossed}) and Eq.(\ref{eq:AHE_NCA}), we see that the inclusion of the $X$ and $\Psi$ diagrams cancels  most of the contribution from the non-crossing diagram in the leading order of $u/v$, as shown in Fig.\ref{fig:AH_conductivity}. This is similar to the case of the 2D massive Dirac model at large energy with Gaussian disorder. However, for the 2D massive Dirac model, the inclusion of the $X$ and $\Psi$ diagram changes the dependence of the total anomalous Hall conductivity on the energy from $\sigma^{\rm NCA}_{xy}\sim m/\epsilon_F$ to $\sigma^{\rm total}_{xy}\sim (m/\epsilon_F)^3$,  which greatly reduces the total anomalous Hall conductivity in the metallic regime $m/\epsilon_F\ll 1$. Whereas for tilted Weyl metals, the contributions of the $X$ and $\Psi$ diagrams have the same dependence on the Fermi energy as the non-crossing diagram  and the cancellation is due to the opposite signs but close values of the coefficients of the two contributions.

\section{Discussions}

The expansion of the response functions of the $X$ and $\Psi$ diagrams to the second order of $\mathbf u$ reveals that the contributions to the AHE from these diagrams  vanish in the second order of $\mathbf u$ for the tilted Weyl metals. The next leading order corrections to the AHE from the $X$ and $\Psi$ diagrams are then $\sim (u/v)^3$. The same is true for the contribution to the AHE from the non-crossing diagram~\cite{Chen2022}. For the type-I Weyl metals with not very large $u/v$,  the anomalous Hall conductivity  in the linear order of $u$ we obtained in this work is then accurate enough.

The contributions to the AHE from the $X$ and $\Psi$ diagrams  do not depend on the disorder strength and scattering rate for Gaussian disorder, and have  the same dependence on the Fermi energy and the tilting of the Weyl metals as the NCA diagram in the leading order. This makes it hard to distinguish the contributions from the two types of diagrams in experiments. However, the  skew scattering contribution comes from consecutive scatterings off two closely located impurities with distance of the order of electron Fermi wavelength~\cite{Ado2015, Ado2016, Ado2017, Levchenko2016}. The impurity density required to observe the AHE due to the $X$ and $\Psi$ diagrams as well as the skew scattering contribution in the NCA diagram is then much higher than that to observe the side jump effects originating from incoherent single impurity scatterings. The self-average of the impurities in the diagrammatic technique indicates an average over all the independent and equivalent subsystems of the size of the phase coherence length $l_\phi$.
To validate the  self-average over the impurities in the calculation of the skew scattering contribution, every independent sub-system of the size of the phase coherence length $l_\phi$ (which is much smaller than the sample size) needs to contain at least one pair of such closely located impurities. We can then estimate the minimum impurity density required in the system to observe the effects of the $X$ and $\Psi$ diagrams as follows. Assume there are $N$ randomly distributed impurities in each independent subsystem of the size of $\l_\phi$. The probability for two randomly chosen impurities in this subsystem to have distance less or equal to the Fermi wavelength $\lambda_F$ is  $\sim(\lambda_F/l_\phi)^3$. Since there are  $N(N-1)/2$ ways to choose a pair of impurities in the subsystem, the total pair number of the rare impurity complexes in the subsystem is $\sim \frac{N(N-1)}{2} (\lambda_F/l_\phi)^3$. This pair number needs to be greater than one to validate the results of the skew scattering contribution, including the $X$ and $\Psi$ diagrams and the non-crossing skew scattering diagrams in Ref.~\cite{Sinitsyn2007, Chen2022}, so we get $N>\sqrt{2}(l_\phi/\lambda_F)^{3/2}$ and the impurity density $n^{sk}_{\rm imp}>\sqrt{2}/(\lambda_F l_\phi)^{3/2}$. (Note that under  this impurity density, the condition $k_F l \gg1$ can still be  satisfied.) As a comparison, the  impurity density required to observe the incoherent single impurity scattering effect, e.g., the side jump contribution, is $n^{sj}_{\rm imp}> 1/(l_\phi)^3$. Since $l_{\phi}> l\gg \lambda_F$, the minimum impurity density to observe the AHE due to skew scatterings is much higher than that of observing the side jump contribution. The same is true for the 2D massive Dirac systems for which $n_{\rm imp}^{sj}> 1/(\l_\phi)^2$ and  $n^{sk}_{\rm imp} > \sqrt{2}/(\lambda_F l_\phi)$.

For the recently studied type-I Weyl metal ${\rm Co_3Sn_2S_2}$ in experiments~\cite{Felser2018, Ding2019}, 
  the topological Chern-Simons term~\cite{Chen2022, Burkov2014, Burkov2015} gives an extra anomalous Hall conductivity $\sigma_H\sim \frac{e^2}{2\pi^2}{\cal K}$ which is proportional to the distance  ${\cal K}$  between the two Weyl nodes. This contribution is independent of the impurity scatterings and constitutes part of the intrinsic AHE.   For  ${\rm Co_3Sn_2S_2}$, the AHE from the Chern-Simons term is one order of magnitude greater than both the contribution from the non-crossing diagram and the crossed diagrams of the low energy effective Hamiltonian with Gaussian disorder~\cite{Fu2021} so it dominates the total AHE in this system. This makes it hard to distinguish the contribution of the disorder in experiments, either due to the NCA diagram or the crossed diagrams. In Ref.~\cite{Felser2018}, all the AHE measured in ${\rm Co_3Sn_2S_2}$ was attributed to the intrinsic one. In Ref.~\cite{Ding2019}, the authors measured the AHE in both the clean and dirty samples, but the difference of the anomalous Hall conductivity in the two samples is only about $10\%$ of the clean case. To better observe the effects of the disorder and the interplay of the non-crossing and crossed diagrams in experiments, one may increase the Fermi level of ${\rm Co_3Sn_2S_2}$ by doping so to enhance the weight of the contribution to the AHE due to both the non-crossing and crossed diagrams since the anomalous Hall conductivity from the Chern-Simons term does not depend on the Fermi energy.

Another way to observe the disorder effects, and  the interplay between the crossed diagrams and the non-crossing diagram
 in ${\rm Co_3Sn_2S_2}$  is by the measurement of the anomalous Nernst effect (ANE)~\cite{Ding2019, Sakai2018, Guin2019} in such system. The ANE only comes from the scatterings on the Fermi surface so the Chern-Simons term has no contribution to the ANE. The ANE is  proportional to the  Fermi surface contribution of the AHE, i,e., $\sigma^I_H$ we studied in Ref.~\cite{Chen2022} and this work, with the ratio $\sim k_B T/\epsilon_F$~\cite{Fu2021}. By measuring the ANE in different disorder conditions, one can tell whether and when the skew scatterings play a role in both the ANE and AHE in the system. Indeed, in Ref.~\cite{Ding2019}, the ANEs in the disordered samples are about three to five times of the clean sample, which makes the effects of the disorder much more discernible in the ANE than in the AHE. The large enhancement of the ANE by the disorder in this experiment agrees qualitatively with our calculation for the NCA diagram in the previous work~\cite{Chen2022} and seems to indicate that the crossed diagrams do not contribute to the ANE in the measured disordered samples based on our calculation of the Gaussian disorder in this work. One possible reason may be that the impurity densities in the disordered samples in the experiment did not reach the required density $n^{sk}_{\rm imp}$ to observe the skew scattering effects. On the other hand,  the real system may also include  disorders more complicated than the Gaussian disorder  considered in this work~\cite{ Ado2017, Offidani2018, Nagaosa2008, Huang2016, Ferreira2016}, which may change the results of the AHE and ANE significantly. For example, it was shown in Ref.\cite{Ado2017} that for the 2D massive Dirac model with smooth disorder, the anomalous Hall conductivity is enhanced by the $X$ and $\Psi$ diagrams instead of being canceled as in the case of Gaussian disorder. Impurities with higher order correlations than the Gaussian disorder may also introduce new contribution to the AHE~\cite{Nagaosa2008}. Besides, the impurities with internal structure may also activate new skew scattering mechanism and change the AHE significantly~\cite{Huang2016, Ferreira2016}. A more complete theoretical study including various realistic disorder is then needed to tell whether the crossed diagrams play a role in such experiments. We will leave this for a future study since  the study of the 3D tilted Weyl metals with these types of disorder is more complicated than the 2D massive Dirac model due to the increased dimensionality. On the other hand, to observe the AHE or ANE due to  the crossed diagrams, more experiments with a more wide range of disorder conditions may also need to be carried out in the future.

\section{Summary}
To sum up, we study the AHE due to the crossed $X$ and $\Psi$ diagrams in Type-I Weyl metals with Gaussian disorder. We show that similar to the 2D massive Dirac model, the contributions from the crossed diagrams cancel a majority part of the contribution from the non-crossing diagram of the low energy effective Hamiltonian. However, the impurity density needed to observe the AHE due to the $X$ and $\Psi$ diagrams is much higher than that of observing the contribution of the non-crossing diagrams with single impurity scatterings.  We estimate the minimum impurity density to observe the AHE due to the $X$ and $\Psi$ diagrams  and discuss the experimental relevance to observe the AHE from such crossed diagrams in the type-I Weyl metals ${\rm Co_3Sn_2S_2}$.

\section{Acknowledgement}
This work is supported by the National Natural Science Foundation of China under Grant No. 11974166.

\begin{widetext}

\appendix
\section{The calculation of $I^X$ and $I^\Psi$}

In this appendix, we show the details of the integration in Eq.(\ref{eq:I_X}) and (\ref{eq:S_Psi1}) -(\ref{eq:S_Psi2}) 
for the $X$ and $\Psi$ diagram.

We first present the integration 
\begin{eqnarray}\label{eq:integration_X}
   I^X&= & {\rm Im}\int_0^{\infty}d Q S^X(Q)  \nonumber\\
   &=& {\rm Im}\int_0^{\infty}Q^4 d Q \  \int_{0}^{\pi}\sin^3\theta_1 d\theta_1 \int_{0}^{\pi}\sin^3\theta_2 d\theta_2 \frac{1}{(v^2Q^2+2v\epsilon_F Q \cos\theta_1-\frac{i}{\tau}\epsilon_F)^2} \frac{1}{v^2Q^2+2v\epsilon_F Q \cos\theta_2+\frac{i}{\tau}\epsilon_F}. \nonumber
\end{eqnarray}

By a change of the variable $x=\cos\theta_1, y=\cos\theta_2$ and denoting $\epsilon_F$ as $\epsilon$ for brevity, we get
\begin{eqnarray}
    I^X
    &=&\int_0^{\infty}Q^4 d Q  \int_{-1}^{1}dx  \int_{-1}^{1}dy (1-x^2)(1-y^2)\nonumber\\
    &&\times
    \{-\frac{\frac{\epsilon}{\tau}}{(v^2Q^2+2v\epsilon Q y)^2+\frac{\epsilon^2}{\tau^2}} 
   \frac{(v^2Q^2+2v\epsilon Q x)^2-\frac{\epsilon^2}{\tau^2}}{[(v^2Q^2+2v\epsilon Q x)^2+\frac{\epsilon^2}{\tau^2}]^2} 
   + \frac{2\frac{\epsilon}{\tau}(v^2Q^2+2v\epsilon Q y)}{(v^2Q^2+2v\epsilon Q y)^2+\frac{\epsilon^2}{\tau^2}}\frac{v^2Q^2+2v\epsilon Q x}{[(v^2Q^2+2v\epsilon Q x)^2+\frac{\epsilon^2}{\tau^2}]^2}\}. \nonumber\\
\end{eqnarray}

We denote the first and second term in the above equation as $I^{X, 1}$ and $I^{X,2}$ respectively and calculate them separately in the following. With the variable substitution $v^2Q^2+2v\epsilon Q y=t,  v^2Q^2+2v\epsilon Q x=s$ and the relationship
\begin{equation}
    \frac{\frac{\epsilon}{\tau}}{(v^2Q^2+2v\epsilon Q y)^2+\frac{\epsilon^2}{\tau^2}}\approx\pi \delta(v^2Q^2+2v\epsilon Q y),
\end{equation}
we get 
\begin{eqnarray}\label{eq:I_X1}
    I^{X,1}
    &=&-\pi\int_0^{\infty}\frac{Q^4}{(2\epsilon vQ)^2}  d Q \int_{v^2Q^2-2v \epsilon Q}^{v^2Q^2+2v \epsilon Q} ds [1-(\frac{s-v^2Q^2}{2v \epsilon Q})^2]
 \frac{s^2-\frac{\epsilon^2}{\tau^2}}{(s^2+\frac{\epsilon^2}{\tau^2})^2}  \int_{v^2Q^2-2v \epsilon Q}^{v^2Q^2+2v \epsilon Q} dt [1-(\frac{t-v^2Q^2}{2v \epsilon Q})^2]\delta(t),\label{eq:I_X1}\\
    I^{X,2}&
    =& \frac{2\epsilon}{\tau}\int_0^{\infty}\frac{Q^4}{(2\epsilon vQ)^2}  d Q   \int_{v^2Q^2-2v \epsilon Q}^{v^2Q^2+2v \epsilon Q} ds [1-(\frac{s-v^2Q^2}{2v \epsilon Q})^2]\frac{s}{(s^2+c^2)^2} \int_{v^2Q^2-2v \epsilon Q}^{v^2Q^2+2v \epsilon Q} dt [1-(\frac{t-v^2Q^2}{2v \epsilon Q})^2]\frac{t}{t^2+ c^2}. \nonumber\\
\end{eqnarray}

We first do the integration of $I^{X,1}$. The integration over $s$ and $t$ for $I^{X,1}$ in Eq.(\ref{eq:I_X1}) can be carried out separately at first. To get a non-vanishing integration over the  $\delta(t)$ factor in Eq.(\ref{eq:I_X1}), $Q$ must be limited to $0<Q<2\epsilon/v$. Denoting $c\equiv\epsilon/\tau$ and integrating out $s$ and $t$, $I^{X, 1}$ becomes 
\begin{eqnarray}
    I^{X,1}&=&\int_0^{2\epsilon/v}d Q \frac{-\pi Q^4 }{(2\epsilon v Q)^2} [1-\frac{v^2Q^2}{4\epsilon^2}]   \nonumber\\
    &&\times\{-(1-\frac{v^2Q^2}{4\epsilon^2}) \frac{s}{s^2+c^2}   +\frac{1}{2\epsilon^2}[c\pi \delta(s) +\frac{1}{2}\ln (s^2+c^2)] 
    -\frac{1}{4\epsilon^2 v^2 Q^2} [ s+c s\delta(s)-2c\arctan(\frac{s}{c}) ]\} |_{s=v^2Q^2-2v \epsilon Q}^{v^2Q^2+2v \epsilon Q}. \nonumber\\
\end{eqnarray}
Neglecting the terms small in $1/\tau$, we get 
\begin{eqnarray}
    I^{X,1}&=&\int_0^{2\epsilon/v}d Q \frac{-\pi Q^4 }{(2\epsilon v Q)^2} [1-\frac{v^2Q^2}{4\epsilon^2}]  \times 
  \{-(1-\frac{v^2Q^2}{4\epsilon^2}) \frac{s}{s^2+c^2}   +\frac{1}{4\epsilon^2}\ln (s^2+c^2)
    -\frac{1}{4\epsilon^2 v^2 Q^2}  s\} |_{s=v^2Q^2-2v \epsilon Q}^{\ \  v^2Q^2+2v \epsilon Q} \nonumber\\
    &=&\frac{\pi}{2\epsilon v^5}[1-\frac{1}{15}(1+8\ln 2)].
\end{eqnarray}

Similarly, after the integration over $s$ and $t$, $I^{X, 2}$ becomes
\begin{eqnarray}
    I^{X, 2}&
    =&2 \int_0^{\infty}\frac{Q^4 }{(2\epsilon vQ)^2}d Q  \left.[ (1-\frac{v^2Q^2}{4\epsilon^2})\frac{1}{2}\ln (t^2+c^2) +\frac{1}{2\epsilon^2}t-\frac{1}{8\epsilon^2 v^2 Q^2 }t^2]\right|_{t=v^2Q^2-2v \epsilon Q}^{\ \ v^2Q^2+2v \epsilon Q}\nonumber\\
    &&\times \left.[-(1-\frac{v^2Q^2}{4\epsilon^2})\frac{\pi}{2}\delta(s)+ \frac{1}{4\epsilon^2}\arctan(\frac{s}{c})]\right|_{s=v^2Q^2-2v \epsilon Q}^{\ \ v^2Q^2+2v \epsilon Q}, \label{eq:I_X_2}
\end{eqnarray}
where we have omitted the terms proportional to $c$ or $s\delta(s)$. The integration over the terms with $\delta(s)$ in Eq.(\ref{eq:I_X_2}) is zero because at $s=0$, $Q=0$ or $1-\frac{v^2Q^2}{4\epsilon^2}=0$, and the terms with $\delta(s)$ in the integrand become zero. For the reason, we only need to consider the terms with $\frac{1}{4\epsilon^2}\arctan(\frac{s}{c})$ after the integration of $s$. Since $c$ is small, 
$\arctan(\frac{s}{c})|_{s=v^2Q^2-2v \epsilon Q}^{\ \ v^2Q^2+2v \epsilon Q}$
 is non-zero only when $v^2Q^2-2v \epsilon Q<0<v^2Q^2+2v \epsilon Q$, i.e., $0<Q<2\epsilon/v$. In this regime, 
 \begin{equation}
\arctan(\frac{s}{c})|_{s=v^2Q^2-2v \epsilon Q}^{\ \ v^2Q^2+2v \epsilon Q}\approx\pi,
\end{equation}
and 
\begin{align*}
    I^{X, 2}&
    =\frac{\pi}{2\epsilon^2}\int_0^{2\epsilon/v}\frac{Q^4 }{(2\epsilon vQ)^2}d Q \times [\frac{1}{2}(1-\frac{v^2Q^2}{4\epsilon^2})\ln\frac{(v^2Q^2+2v \epsilon Q)^2+c^2}{(v^2Q^2-2v \epsilon Q)^2+c^2} +\frac{vQ}{\epsilon} ]\nonumber\\
    &=\frac{\pi}{2\epsilon v^5}[1+\frac{1}{15}(1+8\ln 2)].
\end{align*}

Adding $I^{X, 1}$ and $I^{X, 2}$ together, we get $I^X=\pi/\epsilon_F v^5$ as in Eq.(\ref{eq:I_X}).

\

We next compute $I^{\Psi}\equiv {\rm Im}\int_0^{\infty}d Q S^\Psi(Q)$ for the $\Psi$ diagram. As shown in the main text, we divide $I^{\Psi}$ to two parts $I^{\Psi, 1}$ and $I^{\Psi, 2}$ and  compute them separately. 

From Eq.(\ref{eq:S_Psi_1}), we get
\begin{eqnarray}\label{eq:Psi_1}
  I^{\Psi,1}&=&{\rm Im}  \int_0^{\infty}dQ S^{\Psi,1}(Q) \nonumber\\
  &=&\int_0^{\infty}\frac{Q^4}{4\epsilon^2} d Q  \int_{-1}^{1}dx  \int_{-1}^{1}dy (1-x^2)(1-y^2)\nonumber\\
    &&\times
    [\frac{\frac{\epsilon}{\tau}}{(v^2Q^2-2v\epsilon Q y)^2+\frac{\epsilon^2}{\tau^2}} 
   \frac{(v^2Q^2-2v\epsilon Q x)}{(v^2Q^2-2v\epsilon Q x)^2+\frac{\epsilon^2}{\tau^2}} 
   + \frac{\frac{\epsilon}{\tau}}{(v^2Q^2-2v\epsilon Q x)^2+\frac{\epsilon^2}{\tau^2}}\frac{v^2Q^2-2v\epsilon Q y}{(v^2Q^2-2v\epsilon Q y)^2+\frac{\epsilon^2}{\tau^2}}] \nonumber\\
   &=&\frac{\pi}{2\epsilon^2} \int_0^{\infty}\frac{Q^4}{(2\epsilon vQ)^2} d Q 
    \int_{v^2Q^2-2v \epsilon Q}^{v^2Q^2+2v \epsilon Q} ds [1-(\frac{s-v^2Q^2}{2v \epsilon Q})^2]
 \frac{s}{s^2+\frac{\epsilon^2}{\tau^2}} 
 \int_{v^2Q^2-2v \epsilon Q}^{v^2Q^2+2v \epsilon Q}dt [1-(\frac{t-v^2Q^2}{2v \epsilon Q})^2]\delta(t) \nonumber\\
  &=&\frac{\pi}{2\epsilon^2} \int_0^{2\epsilon/v}d Q \frac{Q^4}{(2\epsilon vQ)^2}  (1-\frac{v^2Q^2}{4\epsilon^2})
 [ (1-\frac{v^2Q^2}{4\epsilon^2})\frac{1}{2}\ln \frac{(v^2Q^2+2v \epsilon Q)^2+c^2}{(v^2Q^2-2v \epsilon Q)^2+c^2  }+\frac{vQ}{\epsilon}] \nonumber\\
 &=&\frac{\pi}{\epsilon v^5}\frac{17+16\ln 2}{105}.
\end{eqnarray}

Similarly, we get 
\begin{eqnarray}
 I^{\Psi,2}&=&-\int_0^{\infty}Q^4 d Q \times {\rm Im} \int_{-1}^{1}dx  \int_{-1}^{1}dy (1-x^2)(1-y^2)[\frac{1}{(v^2Q^2-2v\epsilon Q x+\frac{i}{\tau}\epsilon)^2}\times \frac{1}{v^2Q^2-2v\epsilon Q y+\frac{i}{\tau}\epsilon} ] \nonumber\\
    &=&\int_0^{\infty}Q^4 d Q  \int_{-1}^{1}dx  \int_{-1}^{1}dy (1-x^2)(1-y^2) \nonumber\\
    &&\times
    [\frac{\frac{\epsilon}{\tau}}{(v^2Q^2-2v\epsilon Q y)^2+\frac{\epsilon^2}{\tau^2}} 
   \frac{(v^2Q^2-2v\epsilon Q x)^2-\frac{\epsilon^2}{\tau^2}}{[(v^2Q^2-2v\epsilon Q x)^2+\frac{\epsilon^2}{\tau^2}]^2} 
   + \frac{2\frac{\epsilon}{\tau}(v^2Q^2-2v\epsilon Q y)}{(v^2Q^2-2v\epsilon Q y)^2+\frac{\epsilon^2}{\tau^2}}\frac{v^2Q^2-2v\epsilon Q x}{[(v^2Q^2-2v\epsilon Q x)^2+\frac{\epsilon^2}{\tau^2}]^2}] \nonumber\\
   &=& -I^{X,1}+I^{X,2} \nonumber\\
   &=&\frac{\pi}{\epsilon v^5}\frac{1}{15}(1+8\ln 2).
\end{eqnarray}

\

\end{widetext}


\begin{thebibliography}{99}

\bibitem{Hall1881} E. Hall, Phil. Mag. {\bf 12}, 157 (1881).


\bibitem{Luttinger1954}R. Karplus and J. M. Luttinger, Phys. Rev. {\bf 95}, 1154(1954).

\bibitem{Haldane1988}F. D. Haldane, Phys. Rev. Lett. {\bf 61}, 2015(1988).

\bibitem{Sinitsyn2008}N. A. Sinitsyn, J Phys.: Cond. Matt. {\bf 20}, 023201(2008).

\bibitem{Sinitsyn2007}N. A. Sinitsyn, A. H. MacDonald, T. Jungwirth, V. K. Dugaev, and J. Sinova, Phys. Rev. B {\bf 75}, 045315(2007).


\bibitem{Yang2011}S. Yang, H. Pan, Y. Yao, and Q. Niu, Phys. Rev. B {\bf 83}, 125122(2011).

\bibitem{Streda1982}P. Streda, J. Phys. C {\bf 15}, L717(1982).

\bibitem{MacDonald2006}N. A. Sinitsyn, J. E. Hill, Hongki Min, Jairo Sinova, and
A. H. MacDonald, Phys. Rev. Lett. {\bf 97}, 106804(2006).


\bibitem{Ado2016}I. A. Ado, I. A. Dmitriev, P. M. Ostrovsky, and M. Titov, Phys.Rev. Lett. {\bf 117}, 046601(2016).


\bibitem{Ado2015} I. A. Ado, I. A. Dmitriev, P. M. Ostrovsky, and M. Titov,
Europhys. Lett. {\bf 111}, 37004 (2015).


\bibitem{Ado2017}I. A. Ado, I. A. Dmitriev, P. M. Ostrovsky, and M. Titov, Phys. Rev. B {\bf 96}, 235148(2017).


\bibitem{Levchenko2016}E.J. Konig, P. M. Ostrovsky, M. Dzero, and A. Levchenko, Phys. Rev. B {\bf 94}, 041403(R)(2016).


\bibitem{Inoue2006}J.-I. Inoue, T. Kato, Y. Ishikawa, H. Itoh, G. E. W. Bauer, and L. W. Molenkamp, Phys. Rev. Lett. {\bf 97}, 046604 (2006).


\bibitem{Sinova2007}T. S. Nunner, N. A. Sinitsyn, M. F. Borunda, V. K. Dugaev, A. A. Kovalev, A. Abanov, C. Timm, T. Jungwirth, J. I. Inoue, A. H. MacDonald, and J. Sinova, Phys. Rev. B {\bf 76}, 235312 (2007).

\bibitem{Levchenko2017} E. J. König and A. Levchenko, Phys. Rev. Lett. {\bf 118}, 027001(2017).

\bibitem{Milletari2016}M. Milletari, and A. Ferreira, Phys. Rev. B {\bf 94}, 134202(2016).

\bibitem{Wan2011}X. Wan, A. M. Turner, A. Vishwanath, and S. Y. Savrasov, Phys. Rev. B {\bf 83}, 205101(2011).


\bibitem{Vishwanath2018}N. P. Armitage, E. J. Mele, and A. Vishwanath, Rev. Mod. Phys. {\bf 90}, 015001(2018).

\bibitem{Burkov2014}A. A. Burkov, Phys. Rev. Lett. {\bf 113}, 187202(2014).


\bibitem{Pesin2017}J. F. Steiner, A. V. Andreev, and D. A. Pesin, Phys. Rev. Lett. {\bf 119}, 036601(2017).

\bibitem{Zyuzin2017}Y. Ferreiros, A.A.Zyuzin, and J. H. Bardarson, Phys. Rev. B {\bf 96}, 115202(2017).

\bibitem{Fu2021}M. Papaj, and L. Fu, Phys. Rev. B {\bf 103}, 075424(2021).

\bibitem{Chen2022}J. X. Zhang, Z. Y. Wang, and Wei Chen, Phys. Rev. B {\bf 107}, 125106(2023).

\bibitem{Felser2018}E. Liu, Y. Sun, N. Kumar, L. Muechler, A. Sun, L. Jiao, S. Y. Yang, D. Liu, A. Liang, Q. Xu, J. Kroder, V. Süß, H. Borrmann, C. Shekhar, Z. Wang, C. Xi, W. Wang, W. Schnelle, S. Wirth, Y. Chen, S. T. B. Goennenwein, and C. Felser,  Nat. Phys. {\bf 14}, 1125 (2018).

\bibitem{Ding2019}L. Ding, J. Koo, L. Xu, X. Li, X. Lu, L. Zhao, Q. Wang, Q. Yin, H. Lei, B. Yan, Z. Zhu, and K. Behnia,  Phys. Rev. X {\bf 9}, 041061 (2019)

\bibitem{Wang2018}Wang, Q., Xu, Y., Lou, R., Liu, Z.  Li, M., Y.B,Huang, D.Shen, H. Weng, S. Wang and H. Lei,  Nat. Commun. {\bf 9}, 3681 (2018).


\bibitem{Burkov2015}A. A. Burkov, J Phys.: Cond Matt. {\bf 27}, 113201(2015).

\bibitem{Ye2018}L. Ye, M. Kang, J. Liu, F. von Cube, C. R. Wicker, T. Suzuki, C. Jozwiak, A. Bostwick, E. Rotenberg, D. C. Bell, L. Fu, R. Comin, and J. G. Checkelsky, Nature  {\bf 555}, 638 (2018).


\bibitem{Sakai2018}A. Sakai,  Y. P. Mizuta,  A. A. Nugroho, R. Sihombing, T. Koretsune, M.T. Suzuki {\it et al},  Nat. Phys. {\bf 14}, 1119–1124 (2018).

\bibitem{Guin2019}S. N. Guin,  P. Vir, Y. Zhang,  N. Kumar  and S. J. Watzman {\it et al},  Adv. Mater. {\bf 357}, 1806622 (2019).

\bibitem{Offidani2018}M.Offidani and A. Ferreira, Phys. Rev. Lett. {\bf 121}, 126802(2018).

\bibitem{Nagaosa2008}S. Onoda, N. Sugimoto, and N. Nagaosa, Phys. Rev. B {\bf 77}, 165103(2008).

\bibitem{Huang2016}C. Huang, Y.D. Chong, and M. A. Cazalilla, Phys. Rev. B {\bf 94}, 085414(2016).

\bibitem{Ferreira2016}M. Milletari and A. Ferreira, Phys. Rev. B {\bf 94}, 201402(R)(2016).


\end{thebibliography}
\end{document}